\documentclass[aps,pra,notitlepage,twocolumn,superscriptaddress]{revtex4-1}

\usepackage{graphicx}
\usepackage{amssymb}
\usepackage{amsmath}
\usepackage{ifthen}
\usepackage{braket}
\usepackage{xcolor}
\usepackage{bm}
\usepackage{bbm}
\usepackage{siunitx}
\usepackage{float}
\usepackage{dcolumn}
\usepackage{xcolor}

\usepackage{hyperref}

\newcolumntype{d}[1]{D{.}{.}{#1}}

\usepackage{dcolumn}
\newcolumntype{d}[1]{D{.}{.}{#1}}
\newcolumntype{.}{D{x}{}{9}}
\newcolumntype{,}{D{x}{}{5}}
\newcolumntype{;}{D{x}{}{19}}

\definecolor{garrosgreen}{rgb}{0.1, 0.4, 0.1}
\definecolor{dartmouthgreen}{rgb}{0.05, 0.5, 0.06}
\definecolor{duelferred}{rgb}{0.7, 0.2, 0.1}
\definecolor{cambridgeblue}{rgb}{0.1, 0.3, 1.0}
\definecolor{oxfordblue}{rgb}{0.05, 0.2, 0.7}

\newcommand{\calH}{\mathcal{H}}
\newcommand{\calL}{\mathcal{L}}

\newcommand{\dd}{\mathrm{d}}
\newcommand{\ii}{\mathrm{i}}
\newcommand{\ee}{\mathrm{e}}

\newcommand{\FTK}{\mathcal{F}}

\newcommand{\Mone}{\mathrm{M1}}
\newcommand{\Mtwo}{\mathrm{M2}}
\newcommand{\FSO}{\mathrm{FSO}}
\newcommand{\FSS}{\mathrm{FSS}}
\newcommand{\FSSone}{\mathrm{FSS1}}
\newcommand{\FSStwo}{\mathrm{FSS2}}
\newcommand{\SO}{\mathrm{SO}}
\newcommand{\CSO}{\mathrm{CSO}}

\newcommand{\OC}{\mathrm{OC}}
\newcommand{\tildeH}{\mathcal{H}}
\newcommand{\meV}{\mathrm{meV}}

\newcommand{\eVP}{\mathrm{eVP}}

\newcommand{\crr}{\nonumber \\}

\newlength{\minuslength}
\definecolor{light}{gray}{0.90}
\definecolor{darker}{gray}{0.50}
\definecolor{dark}{gray}{0.30}

\begin{document}

\title{Relativistic and Recoil Corrections to Light--Fermion \\
Vacuum Polarization for Bound Systems
of Spin-0, Spin-1/2, and Spin-1 Particles}

\author{Gregory S. Adkins}
\affiliation{Department of Physics and Astronomy, Franklin \& Marshall College,
Lancaster, Pennsylvania 17604, USA}

\author{Ulrich D. Jentschura}
\affiliation{Department of Physics and LAMOR, Missouri University of Science and
Technology, Rolla, Missouri 65409, USA}

\begin{abstract}
In bound systems whose constituent particles are heavier than 
the electron, the dominant radiative correction to energy
levels is given by light-fermion (electronic) 
vacuum polarization. In consequence,
relativistic and recoil corrections to the one-loop vacuum-polarization
correction are phenomenologically relevant. Here, we generalize 
the treatment, previously accomplished for systems with 
orbiting muons, to bound systems of constituents with more general spins: 
spin-0, spin-1/2, and spin-1. We discuss the application
of our more general expressions to various systems 
of interest, including spinless systems (pionium),
muonic hydrogen and deuterium, and devote special
attention to the excited non-$S$ states
of deuteronium, the bound system of a deuteron and its antiparticle.
The obtained energy corrections are of order $\alpha^5 m_r$,
where $\alpha$ is the fine-structure constant and 
$m_r$ is the reduced mass.
\end{abstract}

\maketitle

%
%
\section{Introduction}
\label{sec1}

In bound systems whose constituent particles
are heavier than the electron, the dominant 
radiative correction to the energy 
of bound states is given by 
electronic vacuum polarization (eVP). 
Examples include muonic bound systems,
including muonic hydrogen and deuterium~\cite{PoEtAl2010,PoEtAl2016,PaEtAl2024},
pionium (the bound $\pi^+ \, \pi^-$ system,
see Refs.~\cite{KoRa2000,GaLyRuGa2001,JeSoIn2002,%
Sc2004plb,Sc2004epjc,Sc2005,GaLyRu2008,GaLyRu2009,DI2005,DI2015}),
pionic hydrogen (the bound $\pi^- \, p$ system,
see Ref.~\cite{pionicH,LyRu2000}),
pionic deuterium~\cite{HaEtAl1998,MeRaRu2005},
kaonic hydrogen~\cite{BeEtAl2005},
kaonic deuterium~\cite{MeRaRu2006,DoMe2011},
antiprotonic deuterium (the bound $\bar p \, d$ system, see 
Refs.~\cite{WyGrNi1985,YaKoKoSu2008,LaCa2021,DuLaCa2023corr,DuLaDE2023,BaEtAl2025}),
and, of course, protonium,
the bound system of a proton and its antiparticle~\cite{DoRi1980,Ba1989,CaRiWy1992,KlBaMaRi2002}.
Recently, the bound system of a deuteron
and its antiparticle has been studied in Ref.~\cite{AdJe2025prr}.
In view of a number of recent investigations 
that could hint at a ``dark-photon'' 
coupling to 
neutrons~\cite{KrEtAl2016,KrEtAl2017woc1,KrEtAl2017woc2,KrEtAl2019,AlEtAl2023,KrEtAl2024}, 
deuteronium could present an ideal candidate for a sensitive 
test of the low-energy sector of the Standard Model.
Yet another example consists of the bound system
of a muon and its antiparticle,
known as {\em dimuonium} (Refs.~\cite{JeSoIvKa1997,KaJeIvSo1998,GiEtAl1998})
or {\em true muonium} (Refs.~\cite{HuMa1971,BrLe2009}).


Let us briefly review the 
hierarchy and scaling of the 
leading terms that contribute to the 
spectrum of the heavy bound systems, before 
discussing the relativistic-recoil correction
to eVP, which is the effect of interest for the
current study. In all the mentioned bound systems,
the leading term 
is the Schr\"{o}dinger--Coulomb energy 
of order $(Z\alpha)^2 m_r$, where $\alpha$ 
is the fine-structure constant,
$m_r$ is the reduced mass,
and $Z$ is the nuclear charge number.
In the heavy systems,
the eVP correction to the bound-state
enters at the order $\alpha \, (Z\alpha)^2 m_r$,
({\em i.e.}, at relative order $\alpha$).
While the dominant eVP corrections
depend on the orbital angular-momentum 
quantum number, they are spin-independent in regard
to the constituent particles of the bound system.

In the heavy systems,
while the leading one-loop eVP enters
at order $\alpha^3 m_r$, two-loop vacuum-polarization 
effects (both irreducible as well as reducible 
diagrams) give corrections of order
$\alpha^4 m_r$~\cite{KaSa1955,Sc1989vol3,Bl1972,Hu1976,BoRi1982,LaJe2024,AdJe2025prr}.
Three-loop vacuum-polarization terms
are of order $\alpha^5 m_r$
(see Refs.~\cite{KiLi1983,BrKaTa1992,BaBr1995,KiNo1999prl,%
KiNo1999prd,IvKoKa2009,On2022,AdJe2025threeloop,FoNeTa2025}).
All of the above mentioned corrections are
still spin-independent in regard to the 
constituent particles of the bound system.

The spin-dependence of the bound-state spectrum 
is first encountered when one considers relativistic 
(Breit) corrections to the 
bound spectrum, which enter at order $(Z \alpha)^4 \, m_r$
(see Chap.~11 of Ref.~\cite{JeAd2022book}).
For the kinetic relativistic corrections,
the scaling of the corrections 
is seen easily, in view of the fact that the atomic 
momentum is of order $p \sim Z \alpha m_r$.
Hence, the relativistic kinetic-energy correction,
given by $- \vec p^{\,4}/(8 m^3)$, 
is suppressed relative to the leading nonrelativistic
term (which reads as $\vec p^{\,2}/(2 m)$), 
by two orders of $Z\alpha$.
Spin-orbit and spin-spin interactions also lead to 
corrections of order $(Z \alpha)^4 \, m_r$.
The relativistic Breit corrections are mediated 
through the exchange of virtual photons,
which constitute tree-level Feynman diagrams
(see Chap.~10 of Ref.~\cite{JeAd2022book}).

In the current investigation, we are concerned
with a conceptually particularly interesting effect
that combines relativistic and recoil effects and loop diagrams, {\em i.e.},
the analysis of relativistic and spin-dependent corrections 
to the leading eVP effect in heavy systems~\cite{Pa1996mu,Je2011pra,%
Bo2011preprintv7,Bo2012,VePa2004,KaIvKo2012,KaIvKa2013,KaKoShIv2017,AdJe2024gauge}.
One easily realizes that the relativistic and 
recoil corrections to the one-loop 
eVP effect enter at order $\alpha (Z\alpha)^4 m_r$,
{\em i.e.}, two orders of $\alpha$ 
higher than the leading vacuum-polarization effect.
The evaluation of the latter terms, 
for muonic hydrogen, has a somewhat 
interesting history, 
as evident from Refs.~\cite{Pa1996mu,Je2011pra,%
Bo2011preprintv7,Bo2012,VePa2004,KaIvKo2012,KaIvKa2013,KaKoShIv2017,AdJe2024gauge}.
For spin-$1/2$ bound systems,
gauge invariance questions are 
nontrivial and have been analyzed recently~\cite{KaIvKa2013,AdJe2024gauge}.
In fact, in order to treat the problem in the 
computationally most effective way,
one needs to use a specific gauge,
which is neither the Feynman nor the Coulomb
gauge, but an optimized gauge 
in which the time-time component of the 
photon propagator is free from any 
frequency dependence and remains static,
despite the fact that a massive photon 
(corresponding to the integrand of the Uehling
potential, see Ref.~\cite{Ue1935}) is being exchanged.
This gauge has been originally introduced in 
Ref.~\cite{Pa1996mu} and recently generalized 
to higher orders in Ref.~\cite{AdJe2024gauge}.

Our aim here is to derive
general expressions for the relativistic recoil
correction to eVP, valid for any bound 
systems with constituent spin-0, spin-$1/2$,
and spin-$1$ particles.
We shall devote special attention to 
pionium, muonic hydrogen and muonic deuterium,
for both fine-structure and hyperfine-structure 
effects (see Sec.~\ref{sec5}),
and we present a numerical evaluation for
the bound system of a deuteron and its antiparticle,
which we refer to as 
deuteronium~\cite{AdJe2025prr}
(see Sec.~\ref{sec6}).
The detailed investigation on
deuteronium, which is a bound system of 
spin-1 particles, constitutes the most complicated 
spin structure among the bound systems
covered in the current investigation,
and is more complex than
encountered for bound systems involving electrons~\cite{JeAd2022book}.
Because the bound deuteron-antideuteron
system is of particular interest for the possible detection of 
New Physics~\cite{AdJe2025prl,AdJe2025prr,AdJe2025trueman}
and for measuring the tensor polarizability of the 
deuteron~\cite{AdJe2025prl},
it is indicated to generalize the 
treatment outlined in Refs.~\cite{Pa1996mu,Je2011pra,%
Bo2011preprintv7,Bo2012,VePa2004,KaIvKo2012,KaIvKa2013,KaKoShIv2017,AdJe2024gauge}
for bound systems of spin-$1/2$ particles,
to the spin-1 case.

We take the following steps.
In Sec.~\ref{sec2}, we investigate
the Nonrelativistic Quantum Electrodynamic
(NRQED) Lagrangian for particles with unit spin, 
and derive the corresponding 
Feynman rules. In Sec.~\ref{sec3},
using the interaction kernels derived 
from the Lagrangian, we calculate
the interaction Hamiltonian for particles
of spin-1 and compare our result 
to previous investigations recorded
in the literature~\cite{ZaPa2010,ZaPaPa2022}.
In order to calculate
the relativistic and recoil corrections
to the vacuum-polarization corrected photon
propagator for spin-1 particles,
we use the vacuum-polarization corrected
photon propagator in the optimized
gauge~\cite{AdJe2024gauge}
(see Sec.~\ref{sec4}).  In Sec.~\ref{sec5} we show that our expression
for the vacuum-polarization corrected relativistic and recoil
contributions can be used to describe the corrections for 
bound states of particles with any combination of spins 0, 1/2, and 1.
Numerical results for the bound system
of a deuteron and its antiparticle (deuteronium)
are presented in Sec.~\ref{sec6}.
Conclusions are reserved for Sec.~\ref{sec7}.

Natural units with $\hbar = c = \epsilon_0=1$ 
are used throughout this investigation.  Also,
we use the usual conventions $\mu,\nu = 0,1,2,3$
for space-time indices, $i,j = 1,2,3$ for spatial
indices, and West--Coast conventions for the 
metric $g_{\mu\nu} = {\rm diag}(1,-1,-1,-1)$.  Values
for the fundamental constants were taken from
CODATA 2022 \cite{MoNeTaTi2025}.

%
%
\section{NRQED Lagrangian and Feynman Rules}
\label{sec2}

In order to find the eVP correction to the Breit Hamiltonian we 
shall start from a one-particle Hamiltonian (for spin-1 particles),
and then derive the Breit interaction Hamiltonian from the interaction
kernels describing the exchange of a massive photon,
the photon mass being integrated as a spectral parameter of vacuum polarization.
For the discussion of the relativistic and recoil
corrections to eVP, we use the photon propagator in 
an optimized Coulomb gauge~\cite{AdJe2024gauge}.
Formally, the photon propagator in the optimized Coulomb gauge
describes the exchange of a photon of mass 
\begin{equation}
\label{deflambda}
\lambda = \frac{2 m_e}{\sqrt{1-v^2}} \,,
\end{equation}
where we use $v \in (0,1)$ as the spectral parameter
of vacuum polarization~\cite{AdJe2024gauge}.
The first-order correction to the
photon propagator in the optimized Coulomb (OC) 
gauge~\cite{AdJe2024gauge} reads as follows,
\begin{equation}
\label{DOC1}
D^{\OC{}:1}_{\mu\nu}(k) = \frac{\alpha}{\pi} \int_0^1 \dd v \, f_1(v)
\begin{pmatrix} \frac{1}{\vec k^2+\lambda^2} & 0 \\ 0 &
\frac{\delta^{i j} - \frac{ k^i k^j}{\vec k^2+\lambda^2} }{k^2-\lambda^2}
\end{pmatrix} \,.
\end{equation}
For our purposes, it is sufficient to 
use the instantaneous approximation $k^2-\lambda^2 \rightarrow
-(\vec k\,^2+\lambda^2)$ for the spatial part.  
The one-loop spectral function is 
\begin{equation}
\label{f1v}
f_1(v) = \frac{v^2(1-\frac{v^2}{3})}{1-v^2} \, .
\end{equation}
For the derivation of the Breit Hamiltonian for 
spin-1 particle, we use the plain Coulomb-gauge photon propagator,
which we recall as follows (see Chap.~9 of Ref.~\cite{JeAd2022book}),
\begin{equation}
\label{DC}
D^C_{\mu\nu}(k) = \begin{pmatrix} \frac{1}{\vec k^2} & 0 \\ 0 & 
\frac{1}{k^2} \left ( \delta^{i j} - \frac{\hat k^i \hat k^j}{\vec k^2} \right ) 
\end{pmatrix} \,.
\end{equation}
We employ the instantaneous approximation $k^2 \rightarrow -\vec k\,^2$.

We also need the low-order Feynman rules for the spin-1 version of NRQED.  We
abstract these rules from the spin-1 Hamiltonian given by Zatorski and
Pachucki \cite{ZaPa2010} in their Eq.~(1). 
We convert their expression for
the Hamiltonian of a spin-1 particle interacting with the electromagnetic
field to a Lagrangian density $\calL$ for a particle of 
mass $m$ and charge $q$,
\begin{multline}
\label{interaction_Lag}
{\cal L} = \psi^\dagger \, 
\Bigl( \ii \partial_t  -q A^0 - \frac{\vec \pi\,^2}{2m} + \frac{\vec \pi\,^4}{8 m^3} 
+ \frac{q}{6} r_E^2 \vec \nabla \cdot \vec E \\
+ \frac{q (\tilde g-1)}{4 m^2} \vec S \cdot 
\left( \vec E \times \vec \pi - \vec \pi \times \vec E \right ) \\
+ \frac{q}{2} Q_E (S^i S^j)^{(2)} \partial_j E^i 
+ \frac{q \, \tilde g}{2m} \vec S \cdot \vec B \Bigr) \, \psi \, ,
\end{multline}
where $r_E$ is the root-mean-square (RMS) electric charge radius, 
$Q_E$ is the electric quadrupole
moment, and $\vec S$ is the spin operator.
Furthermore, $\tilde g$ is the magnetic
$g$-factor, which parameterizes the 
magnetic moment $\vec\mu$ as $\vec\mu = \tilde g \, \mu_m \, \vec S$, 
where $\mu_m$ is the mass-scaled magneton 
\begin{equation}
\mu_m = \frac{e \hbar}{2 m} = 
\frac{m_p}{m} \mu_N \,,
\end{equation}
where $m_p$ is the proton mass, and $\mu_N$ is the 
nuclear magneton.  [The relation between the usual $g$-factor defined by
$\vec \mu = g \mu_N \, \vec S$ and $\tilde g$ is $\tilde g = (m/m_p) g$.]
The quadrupole part of the product of two
vectors (which do not necessarily commute) is
\begin{equation}
\label{rank2}
(u^i v^j)^{(2)} \equiv \frac{u^i v^j + u^j v^i}{2} 
- \frac{\delta^{i j}}{3} \vec u \cdot \vec v \, .
\end{equation}
We also use $A^0$ and $\vec A$ for the scalar and vector 
potentials
and the associated electric and 
magnetic fields $\vec E = -
\vec\nabla A^0 - \frac{\partial \vec A}{\partial t}$ and $\vec B = \vec \nabla
\times \vec A$.  The kinetic momentum operator is 
\begin{equation}
\vec \pi = \vec p - q \vec A = -\ii \, \vec \nabla - q  \vec A \,,
\end{equation}
where $q$ is the particle's charge.  The interaction Lagrangian
(\ref{interaction_Lag}) includes all terms involving inverse masses up to
$1/m^2$.  The RMS electric charge radius 
$r_E$ has dimension of length and is measured in units of 
fermi, while the quadrupole moment $Q_E$ is canonically measured
in units of ${\rm fm}^2$ (the elementary charge 
is divided out right from the start, by convention).

We abstract Feynman rules from the 
Lagrangian~\eqref{interaction_Lag} in the usual way.  For derivatives, we replace
$\partial_i$ acting on an incoming spin-1 particle by 
$\ii p^i$, and
$\overleftarrow{\partial_i}$ acting on an outgoing spin-1 particle becomes
$-\ii p'^i$.  The derivative $\partial_\mu$ acting on a photon carrying momentum
$k$ into the vertex becomes $\ii k^\mu = \ii (p'^\mu - p^\mu)$.

\begin{table}
\caption{\label{table1} We extract the Feynman rules for
spin-1 NRQED from the Lagrangian 
given in Eq.~\eqref{interaction_Lag}.
In the calculation of Feynman 
amplitudes, an unpaired spatial index
is contracted with a spatial index of the 
photon propagator.}
\begin{tabular}{l@{\hspace{0.8cm}}l}
\hline
\hline
\rule[-2mm]{0mm}{10mm}
$\dfrac{\ii \vec p\,^4}{8 m^3}$ & 
kinetic energy correction $(K_4)$ \\[3ex]
$-\ii q$ & Coulomb interaction (C) \\[3ex]
$\dfrac{\ii q}{6} r_E^2 \vec k\,^2$ &
Darwin/finite-size interaction (D) \\[3ex]
$\dfrac{q (\tilde g-1)}{2 m^2} \vec S \cdot
( \vec k \times \vec p \, )$ & 
spin-orbit interaction (SO) 
\\[3ex]
$\dfrac{\ii q Q_E}{2} ( S^i S^j )^{(2)} k^i k^j$ & 
quadrupole interaction (Q) 
\\[3ex]
$\dfrac{\ii q}{2m} \left ( p' + p \right )^i$  & 
  convection interaction (conv) 
\\[3ex]
$-\dfrac{q \tilde g}{2m} ( \vec S \times \vec k \, )^i$ & 
 Fermi spin interaction (FS) 
\\[3ex]
\rule[-4mm]{0mm}{6mm}
$- \dfrac{\ii q^2}{m} \delta^{i j}$ & 
  seagull interaction (SG) \\
\hline
\hline
\end{tabular}
\end{table}
 

For the construction of the Feynman rules of 
NRQED involving spin-1 particles, we now take the terms in 
$\ii \cal L$ from Eq.~\eqref{interaction_Lag}
one by one and 
determine the corresponding rules.
We find the Feynman rules 
listed in Table~\ref{table1}.
The kinetic energy interaction acts on a single spin-1 particle line.  The
Coulomb, Darwin, spin-orbit, and quadrupole interactions are for a spin-1
line interacting with a Coulomb photon.  The convection and Fermi spin
interactions are for a spin-1
line interacting with a transverse photon.
Finally, the seagull interaction is for a spin-1 particle line interacting
with two transverse photons at a point.

\begin{figure*}[t!]
\begin{center}
\begin{minipage}{0.99\linewidth}
\begin{center}
\includegraphics[width=0.91\linewidth]{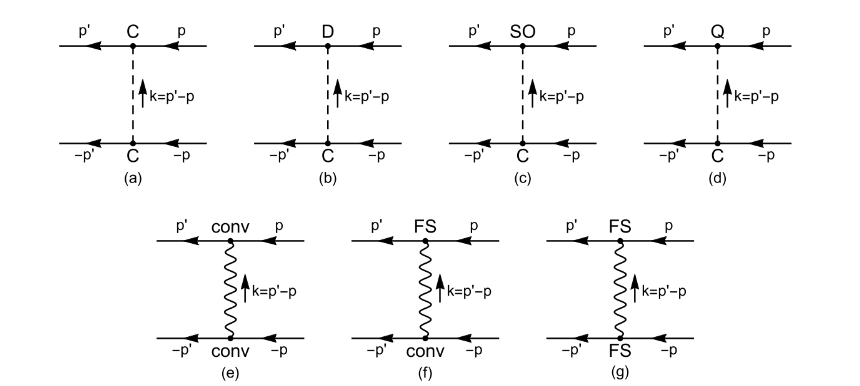}
\end{center}
\caption{\label{Breit} Graphs contributing to the spin-1 Breit Hamiltonian.
Graph (a) represents the basic Coulomb interaction.  Graphs (b)-(d) also
involve the Coulomb photon, and represent the Darwin, spin-orbit, and
quadrupole contributions.  Graphs (e)-(g) contain transverse photons, and
represent the magnetic contribution (e). which involves convection currents (conv), 
the Fermi spin-orbit contribution (f),
and the Fermi spin-spin contribution (g).  Additional contributions with the
vertices on the two spin-1 particles interchanged must also be included for
graphs (b), (c), (d) and (f). For the notation, we refer to Table~\ref{table1}.}
\end{minipage}
\end{center}
\end{figure*}

%
%
\section{Spin-1 Breit Hamiltonian}
\label{sec3}

We will use the Feynman rules
shown in Table~\ref{table1} to construct the
spin-1 Breit Hamiltonian.  We perform this exercise to explicate the
procedure and to confirm the result.  In order to find energy corrections, one
uses the Feynman rules to construct a two-particle to two-particle 
interaction kernel $\delta K$, multiplies
by the imaginary unit, and finds the expectation value in momentum space.
The interaction Hamiltonian is the coordinate-space version of $\ii \delta K$.
As a first example, one considers the simple graph 
given in Fig.~\ref{Breit}(a) consisting
of a Coulomb interaction for each particle connected by a Coulomb photon.  The
energy shift is 

\begin{equation}
\delta E = \ii 
\int \frac{\dd^3 p'}{(2\pi)^3} \, 
\int \frac{\dd^3 p}{(2\pi)^3} \, 
\psi^\dagger(\vec p \, ') (-\ii q_1) \frac{\ii}{\vec k^2} (-\ii q_2) \psi(\vec p \, ) \, .
\end{equation}
Now, we use Fourier transforms to express the wave functions in coordinate space,
\begin{eqnarray}
\delta E &=& (q_1 q_2) \int \frac{\dd^3 p'}{(2\pi)^3}
\int \frac{\dd^3 p}{(2\pi)^3} \, \int \dd^3 x' \, \int \dd^3 x 
\nonumber\\
& & \times
\psi^\dagger(\vec r\, ' ) \, \ee^{\ii \, \vec p\, ' \cdot \vec x\,'} 
\frac{1}{\vec k^2} \ee^{-\ii \, \vec p \cdot \vec x} \, \psi(\vec x\,) 
\nonumber\\
&=& (-4 \pi Z \alpha) 
\int \dd^3 x' \, \int \dd^3 x \, 
\delta^{(3)}(\vec x\, ' - \vec x ) 
\nonumber\\
& & \times \psi^\dagger(\vec x\,') 
\int \frac{\dd^3 k}{(2\pi)^3} 
\frac{\ee^{\ii \vec k \cdot \vec x}}{\vec k^2} 
\psi(\vec x\,) \, .
\end{eqnarray}
It is advantageous to 
use the identity $\exp(\ii \, \vec p\, ' \cdot \vec x\,' \!- \ii \, \vec p \cdot \vec x \, ) =
\exp(\ii \, \vec k \cdot \vec x\, ' + \ii \vec p \cdot (\vec x\, ' - \vec x\, ))$
and to transform the integration measure
$\int \dd^3 p' \, \int \dd^3 p \to \int \dd^3 k \, \int \dd^3 p$.
We assume two spin-1 particles interacting,
one, with a charge $q_1 = \pm |e|$, and 
another, with a charge $q_2 = \mp Z | e |$.
Here, $e$ is the electron charge and $|e|$ is its modulus.
The charge product is $q_1 q_2 = -4 \pi Z\alpha$.
One finds, after a Fourier transformation,
\begin{equation}
\delta E = \int \dd^3 x \, \psi^\dagger(\vec x \, ) 
\left( - \frac{Z \alpha}{r} \right) \psi(\vec x \, ) 
\end{equation}
(where $r \equiv \vert \vec x\, \vert$), so we can identify the corresponding contribution to the Hamiltonian
\begin{equation}
\delta H_C = V_C(r) =  - \frac{Z \alpha}{r} \, ,
\end{equation}
which is the Coulomb potential.  For a kernel contribution with a factor of
$\vec p$, say $\delta K = F^i(\vec k\,) \, p^i$, we can replace the $p^i$ by 
$\ii \partial_i$ acting on $\ee^{-\ii \vec p \cdot \vec x}$.  Then, by partial
integration, the derivative can be moved to act on the wave function $\psi(\vec
x)$ as, now, $p^i = -\ii \partial_i$.  Similarly, for a kernel contribution
$\delta K = p'^i F^i(\vec k\,)$, the $p'^i$ can be replaced by $-\ii \partial'_i$
acting on $\ee^{\ii \vec p\,' \cdot \vec x}$, which can be replaced by 
$\ii \partial'_i$ acting to the left on the wavefunction.  After use of
$\delta^{(3)}(\vec x\,'-\vec x\,)$, another integration by parts allows the
derivative to be replaced by $p^i = -\ii \partial_i$ acting to the right.
Finally, for a kernel $\delta K(\vec p\,', \vec p, \vec k)$, with
$\vec k = \vec p \, ' - \vec p$, all factors of
$\vec p$ can be replaced by operators $\vec p$ acting on the right-hand
wavefunction, while all factors of $\vec p\,'$ can be replaced by the operators
$\vec p$ located just to the right of the left-hand wavefunction and acting to
the right.  In the center is the Fourier transform of the 
$\vec k$-dependence of
$\delta K(\vec p\,', \vec p, \vec k)$.

The Breit Hamiltonian is obtained by transforming kernels 
having order $1/m^2$ (which is tantamount to 
relative order $(Z\alpha)^2$).
These can come from $1/m^2$ interactions times a Coulomb interaction with a
Coulomb propagator, or from products of the $1/m$ interactions multiplied by a
transverse propagator. These graphs are shown in 
Fig.~\ref{Breit}(b)---\ref{Breit}(f). 
We calculate the contributions of the graphs serially.  For the
Darwin contribution given in Fig.~\ref{Breit}(b), one has

\begin{equation}
K^{(4)}_{{\rm D},1} = \left ( \frac{\ii q_1}{6} r_{E1}^2 \vec k^2 \right ) 
\left ( \frac{\ii}{\vec k^2} \right ) (-\ii q_2) 
= - \ii \frac{2 \pi Z \alpha}{3} r_{E1}^2 \,.
\end{equation}
Its contribution to the Breit Hamiltonian,
\begin{multline}
H^{(4)}_{{\rm D},1} = {\FTK} \big [ \ii \delta K_{\rm D} \big ] = 
\frac{2 \pi Z \alpha}{3} \, r_{E1}^2 \, {\FTK} \big [ 1 \big ] \\
= \frac{2 \pi Z \alpha}{3} \, r_{E1}^2 \, \delta^{(3)}(\vec x\,) \, ,
\end{multline}
is obtained by Fourier transform ${\FTK}$, where
\begin{equation}
{\FTK}[ f ] = 
\int \frac{\dd^3 k}{(2\pi)^3} \ee^{\ii \vec k \cdot \vec x } f(\vec k\,) \,.
\end{equation}
A table of useful Fourier transforms
is given in Appendix~\ref{appa}.
The full Darwin contribution includes also the graph
with particles 1 and 2 switched,
\begin{equation}
\label{Breit_Darwin}
H^{(4)}_{\rm D} =
H^{(4)}_{{\rm D},1} + H^{(4)}_{{\rm D},2} = 
\frac{2 \pi Z \alpha}{3} \, (r_{E1}^2 + r_{E2}^2) \, 
\delta^{(3)}(\vec x\,) \,.
\end{equation}
. 

For the spin-orbit contribution of Fig.~\ref{Breit}(c), which involves a 
Coulomb photon, we must remember that any factors of $\vec p\,'$ in
$\delta K(\vec p\,', \vec p, \vec k)$ can be pulled out of the Fourier
transform and placed to its left, while any factors of $\vec p$ can be pulled
out and placed to its right.  The Coulomb spin-orbit kernel coming from 
Fig.~\ref{Breit}(c) reads as follows,
\begin{equation}
K^{(4)}_{\CSO,1} = \left( \frac{q_1 (\tilde g_1-1)}{2 m_1^2} \vec S_1 \cdot 
\left ( \vec k \times \vec p \, \right ) \right ) 
\left ( \frac{\ii}{\vec k^2} \right ) (-\ii q_2) \, .
\end{equation}
Its Breit contribution is
\begin{equation} \label{Breit_spin_orbit}
H^{(4)}_{\CSO,1} = \ii \frac{(-4 \pi Z \alpha) 
(\tilde g_1-1)}{2 m_1^2} \vec S_1 \cdot 
\left ( {\FTK} \left [ \frac{\vec k}{\vec k^2} \right ] \times \vec p \right ) \, ,
\end{equation}
and after the Fourier transformation, one has
\begin{eqnarray}
H^{(4)}_{\CSO,1} &=& \ii \frac{(-4 \pi Z\alpha) (\tilde g_1-1)}{2 m_1^2} 
\vec S_1 \cdot \left ( \frac{\ii \, \vec r}{4 \pi r^3} \times \vec p \right )
\nonumber\\
&=& \frac{Z \alpha (\tilde g_1-1)}{2 m_1^2} \frac{\vec S_1 \cdot \vec L}{r^3} \, .
\end{eqnarray}
Again, we must add the contribution with the identities of particles 1 and 2
interchanged,
\begin{align}
H^{(4)}_{\CSO} =& \; H^{(4)}_{\CSO,1} + H^{(4)}_{\CSO,2} \\
=& \; 
\frac{Z \alpha (\tilde g_1-1)}{2 m_1^2} \frac{\vec S_1 \cdot \vec L}{r^3} +
\frac{Z \alpha (\tilde g_2-1)}{2 m_2^2} \frac{\vec S_2 \cdot \vec L}{r^3} \, .
\end{align}

The quadrupole kernel from Fig.~\ref{Breit}(d) is
\begin{equation}
K^{(4)}_{{\rm Q},1} = \left ( \ii \frac{q_1 Q_{E1}}{2} ( S_1^i S_1^j )^{(2)} \,
k^i k^j \right ) \left ( \frac{\ii}{\vec k^2} \right ) (-\ii q_2) \, .
\end{equation}
The corresponding Hamiltonian contribution is
\begin{eqnarray} 
H^{(4)}_{{\rm Q},1} &=& \frac{(4 \pi Z \alpha) Q_{E1}}{2} 
( S_1^i S_1^j )^{(2)} {\FTK} \left [ \frac{k^i k^j}{\vec k^2} \right ] 
\nonumber\\
&=& - \frac{3 \, Z \alpha \, Q_{E1}}{2 r^3} \,
( S_1^i S_1^j )^{(2)} \, ( \hat x^i \hat x^j )^{(2)} \,.
\end{eqnarray} 
We have used the identity
\begin{align}
{\FTK} \left[ \frac{k^i k^j}{\vec k^2} \right]  =& \;
\frac{1}{3} \delta^{i j} \delta^{(3)}(\vec x\,) 
+ \frac{1}{4 \pi r^3} \left ( \delta^{i j} - 3 \hat x^i \hat x^j \right ) 
\nonumber\\
=& \; 
\frac{1}{3} \delta^{i j} \delta^{(3)}(\vec x\,)
- \frac{3}{4 \pi r^3} \, ( \hat x^i \hat x^j )^{(2)} \,,
\end{align}
with an obvious definition for the rank-2 component
$( \hat x^i \hat x^j )^{(2)}$. Here,
$\delta^{i j}$ times $( S_1^i S_1^j )^{(2)}$ vanishes
because the quadrupole term $( S_1^i S_1^j )^{(2)}$ is traceless
and we used the definition from Eq.~\eqref{rank2}.
After adding the contribution from 
the  intrinsic quadrupole moment of the second particle, one 
obtains
\begin{eqnarray} 
\label{Breit_quadrupole}
H^{(4)}_{{\rm Q}} &=& H^{(4)}_{{\rm Q},1} + H^{(4)}_{{\rm Q},2} 
\nonumber\\
&=& - \frac{3 \, Z \alpha \, Q_{E1}}{2 r^3} \,
( S_1^i S_1^j )^{(2)} \, ( \hat x^i \hat x^j )^{(2)} 
\nonumber\\
&& - \frac{3 \, Z \alpha \, Q_{E2}}{2 r^3} \,
( S_2^i S_2^j )^{(2)} \, ( \hat x^i \hat x^j )^{(2)} \,.
\end{eqnarray}

The remaining contributions involve a transverse photon instead of a Coulomb
photon.  The first is the magnetic interaction of Fig.~\ref{Breit}(e) with two
convection vertices:
\begin{eqnarray} 
\label{Breit_magnetic}
K^{(4)}_{\rm M} &=& \left ( \frac{\ii q_1}{2m_1} 
\left ( p' + p \right )^i \right ) 
\frac{\ii }{-\vec k^2} 
\left ( \delta^{i j} - \frac{k^i k^j}{\vec k^2} \right ) 
\nonumber\\
& & \times \left ( \frac{\ii q_2}{2m_2} \left ( -p' - p \right )^j \right ) 
\nonumber\\
&=&  \frac{-\ii (-4 \pi Z \alpha)}{4 m_1 m_2} 
(2 p')^i \frac{1}{\vec k^2} \left ( \delta^{i j} - 
\frac{k^i k^j}{\vec k^2} \right ) (2 p)^j 
\nonumber\\
&=& \ii \frac{4 \pi Z \alpha}{m_1 m_2} \frac{p'^i 
( \vec k^2 \delta^{i j} - k^i k^j ) p^j}{\vec k^4} \,,
\end{eqnarray} 
where we applied the instantaneous approximation to replace $k^2$ by $-\vec
k^2$ in the denominator and used the transversality of the transverse
propagator to make the replacements $(p'+p)^i = (2p'-k)^i \rightarrow (2 p')^i$
and $(p'+p)^j = (k+2p)^j \rightarrow (2p)^j$. The magnetic Hamiltonian is thus
\begin{align}
H^{(4)}_{\rm M} =& \; - \frac{4 \pi Z \alpha}{m_1 m_2} \, p^i \, 
{\FTK} \left [ \frac{\delta^{i j}}{\vec k^2} - 
\frac{k^i k^j}{\vec k^4} \right ] p^j 
\nonumber\\
=& \; - \frac{4 \pi Z \alpha}{m_1 m_2} \, p^i 
\left \{ \frac{\delta^{i j}}{4 \pi r} - \frac{1}{8 \pi r} 
\left( \delta^{i j} - \hat x^i \hat x^j \right) \right \} p^j 
\nonumber\\
=& \; - \frac{Z \alpha}{2 m_1 m_2} \, p^i \, 
\left ( \frac{\delta^{i j} + \hat x^i \hat x^j}{r} \right ) p^j \, .
\end{align}

The Fermi spin-orbit contribution of Fig.~\ref{Breit}(f) gives a contribution
\begin{eqnarray} 
\label{Breit_Fermi_spin_orbit}
K^{(4)}_{\FSO,1} &=& \left( -\frac{q_1 \tilde g_1}{2m_1} 
( \vec S_1 \times \vec k\, )^i \right ) \left[ -\frac{\ii }{\vec k^2} 
\left ( \delta^{i j} - \frac{k^i k^j}{\vec k^2} \right ) 
\right]
\nonumber\\
& & \times
\left ( \frac{\ii q_2}{2m_2} \left ( -p' - p \right )^j \right ) 
\nonumber\\
&=& \frac{(-4 \pi Z \alpha) \tilde g_1}{4 m_1 m_2} 
( \vec S_1 \times \vec k )^i \frac{1}{\vec k^2} (k+2p)^i 
\nonumber\\
&=& - \frac{2 \pi Z \alpha \tilde g_1}{m_1 m_2} \vec S_1 \cdot 
\left ( \frac{\vec k}{\vec k^2} \times \vec p \right ) \,.
\end{eqnarray}
The corresponding contribution to the Hamiltonian is
\begin{eqnarray}
H^{(4)}_{\FSO,1} &=& -\ii \, \frac{2 \pi Z\alpha \tilde g_1}{m_1 m_2} 
\vec S_1 \cdot \left ( {\FTK} \left [ \frac{\vec k}{\vec k^2} \right ] 
\times \vec p \right ) 
\nonumber\\
&=& -\ii \, \frac{2 \pi Z \alpha \tilde g_1}{m_1 m_2} 
\vec S_1 \cdot \left ( \frac{\ii \vec x}{4 \pi r^3} \times \vec p \right )
\nonumber\\
&=& \frac{Z \alpha \, \tilde g_1}{2 m_1 m_2} \frac{\vec S_1 \cdot \vec L}{r^3} \, .
\end{eqnarray}
With particles 1 and 2 interchanged, one obtains
\begin{eqnarray}
H^{(4)}_{\FSO} &=& H^{(4)}_{\FSO,1} + H^{(4)}_{\FSO,2} 
\nonumber\\
&=& \frac{Z \alpha \, \tilde g_1}{2 m_1 m_2} \frac{\vec S_1 \cdot \vec L}{r^3} 
+ \frac{Z \alpha \, \tilde g_2}{2 m_1 m_2} \frac{\vec S_2 \cdot \vec L}{r^3} \, .
\end{eqnarray}
Finally, the Fermi spin-spin kernel of Fig.~\ref{Breit}(g)  is
\begin{align} 
\label{Breit_FSS}
K^{(4)}_{\FSS} =& \; \left ( -\frac{q_1 \tilde g_1}{2m_1} 
( \vec S_1 \times \vec k )^i \right ) 
\frac{\ii}{-\vec k^2} \left ( \delta^{i j} - 
\frac{k^i k^j}{\vec k^2} \right ) 
\nonumber\\
& \; \times \left ( -\frac{q_2 \tilde g_2}{2m_2} 
( \vec S_2 \times (-\vec k\,) )^j \right ) 
\nonumber\\
=& \; \ii \, \frac{(-4 \pi Z \alpha) \tilde g_1 \tilde g_2}{4 m_1 m_2} \; 
\frac{( \vec S_1 \times \vec k \, ) \cdot 
( \vec S_2 \times \vec k \, )}{\vec k^2 } 
\nonumber\\
=& \; \ii \, \frac{(-4 \pi Z \alpha) 
\tilde g_1 \tilde g_2}{4 m_1 m_2} \; 
\frac{ \vec S_1 \cdot \vec S_2 \, \vec k^2 - 
\vec S_1 \cdot \vec k \, \vec k \cdot \vec S_2}{\vec k^2 } \, .
\end{align} 
The corresponding Hamiltonian contribution is
\begin{eqnarray} 
H^{(4)}_{\FSS} &=& \frac{\pi \, Z \alpha \, \tilde g_1 \tilde g_2}{m_1 m_2} \, 
S_1^i \, \left\{ {\FTK} \left [ \delta^{i j} - \frac{k^i k^j}{\vec k^2}\right ] 
\right\} S_2^j 
\nonumber\\
&=& \frac{2 \pi \, Z \alpha \, \tilde g_1 \tilde g_2}{3 m_1 m_2} \,
\vec S_1 \cdot \vec S_2 \, \delta^{(3)}(\vec x\,)
\nonumber\\
& & + \frac{3 \, Z \alpha \, \tilde g_1 \tilde g_2}{4 m_1 m_2} 
\frac{S_1^i S_2^j (\hat x^i \hat x^j )^{(2)}}{r^3} \, .
\end{eqnarray} 
The complete Breit Hamiltonian, 
which we denote by $H^{(4)}$ because
the terms are of order $(Z\alpha)^4$,
including all interactions on both particles, is
\begin{equation}
\label{H4}
H^{(4)} = H^{(4)}_{\rm K} + H^{(4)}_{\rm M} + H^{(4)}_{\SO} 
+ H^{(4)}_{\FSS} + H^{(4)}_{\rm Q} + H^{(4)}_{\rm D} \,,
\end{equation}
where
\begin{equation}
\label{H4suppl}
H^{(4)}_{\SO} = H^{(4)}_{\CSO} + H^{(4)}_{\FSO} \,.
\end{equation}
In writing Eq.~\eqref{H4}, we have 
adopted the following sequence:
kinetic, magnetic, spin-dependent, quadrupole, and Darwin (finite size).
For convenience, we list the individual terms,
\begin{subequations}
\label{treeH4}
\begin{eqnarray}
\label{treeH4K}
H^{(4)}_K &=& -\frac{p^4}{8 m_1^3} - \frac{p^4}{8 m_2^3} \, , \\[3pt]
H^{(4)}_{\rm M} &=& - \frac{Z \alpha}{2 m_1 m_2} \, p^i \, 
\left ( \frac{\delta^{i j} + \hat x^i \hat x^j}{r} \right ) p^j \, , \\[3pt]
\label{treeH4SO}
H^{(4)}_{\SO} &=& \frac{Z\alpha}{2} 
\left( \frac{\tilde g_1-1}{m_1^2} 
+ \frac{\tilde g_1}{m_1 m_2} \right)
\frac{\vec S_1 \cdot \vec L}{r^3} 
\nonumber\\[3pt]
& & + \frac{Z\alpha}{2}
\left( \frac{\tilde g_2-1}{m_2^2}
+ \frac{\tilde g_2}{m_1 m_2} \right)
\frac{\vec S_2 \cdot \vec L}{r^3} \,, \\[3pt] 
\label{treeH4FSS}
H^{(4)}_{\FSS} &=& \frac{2 \pi \, Z \alpha \, 
\tilde g_1 \tilde g_2}{3 m_1 m_2} \,
\vec S_1 \cdot \vec S_2 \, \delta^{3}(\vec x\,)
\nonumber\\[3pt]
& & + \frac{3 \, Z \alpha \, \tilde g_1 \tilde g_2}{4 m_1 m_2} 
\frac{S_1^i S_2^j \, ( \hat x^i \hat x^j )^{(2)}}{r^3} \,,
\\[3pt]
\label{treeH4Q}
H^{(4)}_{\rm Q} &=& - \frac{3 \, Z \alpha}{2 r^3}  \biggl\{ Q_{E1} 
( S_1^i S_1^j )^{(2)} 
\nonumber\\[3pt]
& & + Q_{E2} ( S_2^i S_2^j )^{(2)} \biggr\} \,
( \hat x^i \hat x^j )^{(2)} \, , \\[3pt]
\label{treeH4D}
H^{(4)}_{\rm D} &=& \frac{2 \pi}{3} \, Z\alpha \, \left ( r_{E1}^2 + r_{E2}^2 \right ) 
\delta^{3}(\vec x\,) \,.
\end{eqnarray}
\end{subequations}
When applied to deuteronium, our result here agrees with the 
Breit Hamiltonian used in Refs.~\cite{AdJe2025prl,AdJe2025prr}.
The two radius terms $r_{E1}^2$ and $r_{E2}^2$
are in need of an interpretation. Namely,
as it has been explained in Ref.~\cite{PaKa1995,Je2011radii},
for spin-$1/2$ particles, 
the Darwin--Foldy term is inherently included in the 
definitions of the radii $r_{E1}^2$ and $r_{E2}^2$
of the two constituent particles.
So, if $r_{C1}^2$ and $r_{C2}^2$ are the 
charge radii of the two particles,
then 
\begin{subequations}
\label{DFterm}
\begin{align}
r_{E1}^2 \to & \; \frac{3 \delta_I}{4 m_1^2} + r_{C1}^2 \,,
\\
r_{E2}^2 \to & \; \frac{3 \delta_I}{4 m_2^2} + r_{C2}^2 \,,
\end{align}
\end{subequations}
where $\delta_I$ is equal to one
for particles with half-integer spin, and equal to zero
for particles with integer spin.

%
%
\section{Relativistic and Recoil Corrections to eVP}
\label{sec4}

The relativistic and recoil corrections to the one-loop eVP (Uehling) potential
are obtained from the one-loop VP corrected OC-gauge photon propagator of
Eq.~(\ref{DOC1}) in the same way as the Breit Hamiltonian was obtained
from the uncorrected Coulomb gauge propagator. In the instantaneous limit needed
here, this one-loop propagator becomes, 
according to Eq.~\eqref{DOC1},
\begin{equation}
\label{DOC1nonretarded}
D^{\OC:1}_{i j}(k) \rightarrow 
\frac{\alpha}{\pi} \int_0^1 \dd v \, f_1(v) 
\begin{pmatrix} \frac{1}{\vec k^2+\lambda^2} & 0 \\ 0 & 
-\frac{\delta^{i j} - \frac{ k^i k^j}{\vec k^2+\lambda^2} }%
{\vec k^2+\lambda^2} \end{pmatrix} \, ,
\end{equation}
where $f_1$ has been defined in Eq.~\eqref{f1v}.

The modification required for the diagrams of Fig.~\ref{Breit} 
can be represented as
\begin{equation}
\label{trafo45}
H^{(5)}_{\eVP} = \frac{\alpha}{\pi} 
\int_0^1 \dd v \, f_1(v) \, \tildeH^{(4)} \,,
\end{equation}
where $\tildeH^{(4)}$ is obtained
via the same matching procedure as $H^{(4)}$,
but with the Coulomb-gauge photon propagator
replaced with the optimized-gauge
propagator from Eq.~\eqref{DOC1},
\begin{equation}
\tildeH^{(4)}= H^{(4)} \bigg 
\vert_{D^C \rightarrow D^{\OC:1}} \,.
\end{equation}
For terms involving the Coulomb photon, the substitution is easy:
\begin{equation}
\frac{1}{\vec k^2} \rightarrow \frac{1}{\vec k^2+\lambda^2} \, .
\end{equation}
For terms involving the transverse photon, 
a little more work is required as the numerator algebra changes, 
since the substitution is
\begin{equation}
\frac{1}{\vec k^2} \left ( \delta^{i j} - \frac{k^i k^j}{\vec k^2} \right ) 
\rightarrow 
\frac{1}{\vec k^2+\lambda^2} 
\left ( \delta^{i j} - \frac{k^i k^j}{\vec k^2+\lambda^2} \right ) \, .
\end{equation}
For example, the Coulomb potential $H_C = -\frac{Z\alpha}{r}$ becomes
\begin{equation}
\tildeH_U =  {\FTK} \left [ \frac{-4 \pi Z \alpha}{\vec k^2+\lambda^2} \right ] \crr
= (-4 \pi Z \alpha) \frac{\ee^{-\lambda r}}{4 \pi r} \crr
=  \left ( \frac{- Z \alpha}{r} \right ) \ee^{-\lambda r} \, ,
\end{equation}
so that we can easily verify the modification
of the Coulomb potential to be equal to 
\begin{equation}
{\cal V}_U = \frac{\alpha}{\pi} \int_0^1 dv \, f_1(v) \, 
\calH_U = \frac{\alpha}{\pi} 
\int_0^1 dv \, f_1(v) \left ( \frac{- Z \alpha}{r} \right ) \ee^{-\lambda r} \,,
\end{equation}
which is the Uehling potential.

The Coulomb-photon terms are the Darwin (finite-size)
contribution [see Fig.~\eqref{Breit}(b) and
Eq.~\eqref{Breit_Darwin}], 
\begin{eqnarray}
\tildeH^{(4)}_{{\rm D},1} &=& \frac{2 \pi \, Z \alpha}{3} r_{E1}^2 \;
{\FTK}\left[ \frac{\vec k^2}{\vec k^2+\lambda^2} \right ] 
\nonumber\\
&=& \frac{2 \pi \, Z \alpha}{3} r_{E1}^2 
\left \{ \delta^{(3)}(\vec x\,) - \lambda^2 
\frac{\ee^{-\lambda r}}{4 \pi r} \right \}
\end{eqnarray}
(along with the corresponding contribution with the particles interchanged),
the Coulomb spin-orbit term of Fig.~\ref{Breit}(c) and 
Eq.~(\ref{Breit_spin_orbit}),
\begin{eqnarray}
\tildeH^{(4)}_{\CSO,1} &=& 
\ii \frac{(-4 \pi \, Z \alpha) (\tilde g_1-1)}{2 m_1^2} 
\vec S_1 \cdot 
\left ( {\FTK} \left [ \frac{\vec k}{\vec k^2+\lambda^2} 
\right ] \times \vec p \right )
\nonumber\\
&=& \ii \frac{(-4 \pi \, Z \alpha) (\tilde g_1-1)}{2 m_1^2} 
\vec S_1 \cdot 
\left( \frac{\ee^{-\lambda r}}{4 \pi r^3} (1+\lambda r) \, 
\ii \, \vec r \times \vec p \right) 
\crr
&=& \frac{Z \alpha \, (\tilde g_1-1)}{2 m_1^2} 
\frac{\ee^{-\lambda r}}{r^3} (1+\lambda r) \,
\vec S_1 \cdot \vec L \, ,
\end{eqnarray}
with an obvious addition of $\tildeH^{(4)}_{\CSO,2}$
where the particles are interchanged,
\begin{multline}
\label{H4SOK}
\tildeH^{(4)}_{\CSO} =
\frac{Z \alpha \, (\tilde g_1-1)}{2 m_1^2}
\frac{\ee^{-\lambda r}}{r^3} (1+\lambda r) \,
\vec S_1 \cdot \vec L 
\\
+ \frac{Z \alpha \, (\tilde g_2-1)}{2 m_2^2}
\frac{\ee^{-\lambda r}}{r^3} (1+\lambda r) \,
\vec S_2 \cdot \vec L \,,
\end{multline}
and the quadrupole term of Fig.~\ref{Breit}(d) and 
Eq.~\eqref{Breit_quadrupole},
\begin{multline}
\tildeH^{(4)}_{Q,1} = \frac{(4 \pi \, Z \alpha) Q_{E1}}{2} 
( S_1^i S_1^j )^{(2)} \,
{\FTK} \left [ \frac{k^i k^j}{\vec k^2+\lambda^2} \right ] 
= - \frac32 Z \alpha 
\\[0.1133ex]
\times 
Q_{E1}
\frac{\ee^{-\lambda r}}{r^3} \left ( 1+\lambda r + 
\frac{\lambda^2 r^2}{3} \right ) 
( S_1^i S_1^j )^{(2)} \,
( \hat x^i \hat x^j )^{(2)} \, .
\end{multline}
After adding the contribution from the quadrupole moment
of the second particle, one obtains
\begin{multline}
\tildeH^{(4)}_{Q} = \tildeH^{(4)}_{Q,1} + \tildeH^{(4)}_{Q,2} 
= - \frac{3 Z \alpha}{2}
\frac{\ee^{-\lambda r}}{r^3}
\left ( 1+\lambda r + \frac{\lambda^2 r^2}{3} \right )
\\
\times \left\{
Q_{E1} \, ( S_1^i S_1^j )^{(2)} \,
( \hat x^i \hat x^j )^{(2)} +
Q_{E2} \, ( S_2^i S_2^j )^{(2)} \,
( \hat x^i \hat x^j )^{(2)} \right\} \,.
\end{multline}

The transverse photon terms require a bit more work. 
The magnetic interaction
kernel of Fig.~\ref{Breit}(e) and 
Eq.~\eqref{Breit_magnetic},
\begin{align}
{\cal K}^{(4)}_{\rm M} =& \;
\left ( \frac{\ii q_1}{2m_1} \left ( p' + p \right )^i \right ) 
\frac{\ii}{-(\vec k^2+\lambda^2)} 
\nonumber\\
& \; \times \left ( \delta^{i j} - \frac{k^i k^j}{\vec k^2+\lambda^2} \right ) 
\left ( \frac{\ii q_2}{2m_2} \left ( -p' - p \right )^j \right ) 
\nonumber\\
=& \; \ii \, \frac{4 \pi \, Z \alpha}{4 m_1 m_2} 
(2 p'-k)^i \frac{(\vec k^2+\lambda^2) \delta^{i j} - 
k^i k^j}{(\vec k^2+\lambda^2)^2} (k+2p)^j \, ,
\end{align}
gives rise to a Hamiltonian
\begin{eqnarray}
\tildeH^{(4)}_{\rm M} &=& -\frac{4 \pi Z \alpha}{4 m_1 m_2} 
{\FTK} \left [ 4 p^i \frac{(\vec k^2+\lambda^2) \delta^{i j} - 
k^i k^j}{(\vec k^2+\lambda^2)^2} p^j \right.
\nonumber\\
& & \left. + \frac{(2 p^i) \lambda^2 k^i}{(\vec k^2+\lambda^2)^2} - 
\frac{\lambda^2 k^j (2 p^j)}{(\vec k^2+\lambda^2)^2} - 
\frac{\lambda^2 \vec k^2}{(\vec k^2+\lambda^2)^2} \right ] 
\nonumber\\
&=& - \frac{Z \alpha}{2 m_1 m_2} p^i 
\frac{\ee^{-\lambda r}}{r} 
\Big [ \delta^{i j} + (1+\lambda r) \hat x^i \hat x^j  \Big ] p^j 
\nonumber\\
& & - \frac{Z \alpha}{4 m_1 m_2} \frac{\lambda^2 \, \ee^{-\lambda r}}{r} 
\left ( 1 - \frac{\lambda r}{2} \right )
\end{eqnarray}
after use of the identity
\begin{align}
\ii p^i \left ( \ee^{-\lambda r} \hat x^i \right ) - 
\left ( \ee^{-\lambda r} \hat x^i \right ) \ii p^i = & \;
\left [ \partial_i , \ee^{-\lambda r} \hat x^i \right ] 
\nonumber\\
=& \; \frac{2}{r} \ee^{-\lambda r} \left ( 1 - \frac{\lambda r}{2} \right ) \, .
\end{align}
The Fermi spin-orbit term of Fig.~\ref{Breit}(f) 
and Eq.~\eqref{Breit_Fermi_spin_orbit} is
\begin{eqnarray}
{\cal K}^{(4)}_{\FSO,1} &=& \left ( -\frac{q_1 \tilde g_1}{2m_1} 
( \vec S_1 \times \vec k )^i \right ) 
\frac{\ii}{-(\vec k^2+\lambda^2)} 
\nonumber\\
& & \times 
\left ( \delta^{i j} - \frac{k^i k^j}{(\vec k^2+\lambda^2)} \right ) 
\left ( \frac{\ii q_2}{2m_2} ( -k^j - 2p^j ) \right ) \crr
&=& - \frac{4 \pi Z \alpha \tilde g_1}{4 m_1 m_2} 
\frac{ ( \vec S_1 \times \vec k ) \cdot 
(2 \vec p )}{(\vec k^2+\lambda^2)} \, ,
\end{eqnarray}
so that
\begin{eqnarray}
\tildeH^{(4)}_{\FSO,1} &=& -\ii 
\frac{4 \pi \, Z \alpha \, \tilde g_1}{4 m_1 m_2} 
\left ( \vec S_1 \times 
{\FTK} \left [ \frac{\vec k}{\vec k^2+\lambda^2} \right ] \right ) 
\cdot (2 \vec p ) 
\nonumber\\
&=& -\ii \frac{4 \pi \, Z \alpha \, \tilde g_1}{4 m_1 m_2} 
\left ( \vec S_1 \times \frac{\ee^{-\lambda r}}{4 \pi r^3} (1+\lambda r) \, 
\ii \vec x \right ) \cdot (2 \vec p ) 
\nonumber\\
&=& \frac{Z \alpha \, \tilde g_1}{2 m_1 m_2} 
\frac{\ee^{-\lambda r}}{r^3} (1+\lambda r) \, \vec S_1 \cdot \vec L \, .
\end{eqnarray}
There is an obvious addition for ${\calH}^{(4)}_{\FSO,2}$,
resulting in
\begin{multline}
\tildeH^{(4)}_{\FSO} =
\frac{Z \alpha \, \tilde g_1}{2 m_1 m_2}
\frac{\ee^{-\lambda r}}{r^3} (1+\lambda r) \, \vec S_1 \cdot \vec L 
\\
+ \frac{Z \alpha \, \tilde g_2}{2 m_1 m_2}
\frac{\ee^{-\lambda r}}{r^3} (1+\lambda r) \, \vec S_2 \cdot \vec L \,.
\end{multline}
We summarize the sum of $\tildeH^{(4)}_{\CSO}$ and
$\tildeH^{(4)}_{\FSO}$ as the total spin-orbit term
\begin{equation}
\tildeH^{(4)}_{\SO} = 
\tildeH^{(4)}_{\CSO} + 
\tildeH^{(4)}_{\FSO} \,.
\end{equation}

Finally, the Fermi spin-spin contribution of 
Fig.~\ref{Breit}(g) and Eq.~\eqref{Breit_FSS} is
\begin{eqnarray}
{\cal K}^{(4)}_{\FSS} &=& \left ( -\frac{q_1 \tilde g_1}{2m_1} 
( \vec S_1 \times \vec k )^i \right ) 
\frac{\ii}{-(\vec k^2+\lambda^2)} 
\nonumber\\
& & \times
\left ( \delta^{i j} - \frac{k^i k^j}{(\vec k^2+\lambda^2)} \right ) 
\left ( -\frac{q_2 \tilde g_2}{2m_2} 
( \vec S_2 \times (-\vec k) )^j \right ) 
\nonumber\\
&=& -\ii \frac{4 \pi \, Z \alpha \, 
\tilde g_1 \tilde g_2}{4 m_1 m_2} 
\frac{( \vec S_1 \times \vec k ) \cdot 
( \vec S_2 \times \vec k )}{\vec k^2+\lambda^2} \, ,
\end{eqnarray}
so that
\begin{multline}
\tildeH^{(4)}_{\FSS} = \frac{4 \pi \, Z \alpha \,
\tilde g_1 \tilde g_2}{4 m_1 m_2} \, 
S_1^i S_2 ^j \, 
{\FTK} \left [ \frac{\delta^{i j} \vec k^2 - k^i k^j}{\vec k^2+\lambda^2} \right ] 
\\
= \frac{4 \pi \, Z \alpha \,
\tilde g_1 \tilde g_2}{4 m_1 m_2} \, 
\Big [ \frac{\vec S_1 \cdot \vec S_2}{3} \delta^{i j} + (S_1^i S_2 ^j)^{(2)} \Big ] \, 
{\FTK} \left [ \frac{\delta^{i j} \vec k^2 - k^i k^j}{\vec k^2+\lambda^2} \right ] 
\\
= \frac{2 \pi \, Z \alpha \, \tilde g_1 \tilde g_2}{3 m_1 m_2} 
\vec S_1 \cdot \vec S_2 \, 
\left\{ \delta^{(3)}(\vec r)
- \lambda^2 \frac{\ee^{-\lambda r}}{4 \pi r} \right\}
\\
+ \frac{3 \, Z \alpha \, \tilde g_1 \tilde g_2}{4 m_1 m_2} \, 
\frac{\ee^{-\lambda r}}{r^3} 
\left(1 + \lambda r + \frac{\lambda^2 r^2}{3} \right) \,
( S_1^i S_2^j )^{(2)} \; ( \hat x^i \hat x^j )^{(2)} \,.
\end{multline}

In summary, the Hamiltonian for relativistic-recoil corrections 
to electronic VP is given by
\begin{equation}
\label{calH5}
H^{(5)}_{\eVP} = 
H^{(5)}_{\rm M} + H^{(5)}_{\SO} + H^{(5)}_{\FSS}
+ H^{(5)}_{\rm Q} + H^{(5)}_{\rm D} \,,
\end{equation}
where the $H^{(5)}_{\rm X}$ with ${\rm X} \in \{ {\rm M}, \SO, \FSS, {\rm Q}, {\rm D} \}$
are obtained from the corresponding
$\tildeH^{(4)}_{\rm X}$ 
by the integration over the spectral
function of vacuum polarization, which is given in 
Eq.~\eqref{trafo45}. We list the terms as follows,
\begin{subequations}
\label{calH4}
\begin{align}
\label{calH4M}
\tildeH^{(4)}_{\rm M} =& \; - \frac{Z \alpha}{2 m_1 m_2} 
p^i \frac{\ee^{-\lambda r}}{r} \Big [ \delta^{i j} + 
(1+\lambda r) \hat x^i \hat x^j  \Big ] p^j 
\nonumber\\
& \; - \frac{Z \alpha \, \lambda^2 }{4 m_1 m_2} 
\frac{\ee^{-\lambda r}}{r} 
\left ( 1 - \frac{\lambda r}{2} \right ) \,, 
\\[3pt]
\label{calH4SO}
\tildeH^{(4)}_{\SO} =& \; \frac{Z \alpha \,
\ee^{-\lambda r} }{2 r^3} (1+\lambda r) 
\left\{ 
\left( \frac{\tilde g_1-1}{m_1^2}  + 
\frac{\tilde g_1}{m_1 m_2} \right) 
\vec S_1 \cdot \vec L \right.
\nonumber\\
& \; \left. + 
\left( \frac{\tilde g_2-1}{m_2^2}  + 
\frac{\tilde g_2}{m_1 m_2} \right) 
\vec S_2 \cdot \vec L  \right\} \,, 
\\[3pt]
\label{calH4FSS}
\tildeH^{(4)}_{\FSS} =& \; \frac{2 \pi  Z \alpha \, \tilde g_1 \tilde g_2}{3 m_1 m_2}
\vec S_1 \cdot \vec S_2 \,
\left\{ \delta^{(3)}(\vec x \, )
- \lambda^2 \frac{\ee^{-\lambda r}}{4 \pi r} \right\}
\nonumber\\
& \; + \frac{3  Z \alpha \, \tilde g_1 \tilde g_2}{4 m_1 m_2} \,
\frac{\ee^{-\lambda r}}{r^3}
\left(1 + \lambda r + \frac{\lambda^2 r^2}{3} \right) \,
\nonumber\\
& \; \times ( S_1^i S_2^j )^{(2)}  \; ( \hat x^i \hat x^j )^{(2)} \, ,
\\[3pt]
\label{calH4Q}
\tildeH^{(4)}_{\rm Q} =& \; - \frac{3 Z \alpha}{2} 
\frac{\ee^{-\lambda r}}{r^3} 
\left ( 1+\lambda r + \frac{\lambda^2 r^2}{3} \right ) 
\nonumber\\
& \; \times \left\{ 
Q_{E1} \, ( S_1^i S_1^j )^{(2)} \,
( \hat x^i \hat x^j )^{(2)} \right.
\nonumber\\
& \; \left. + Q_{E2} \, ( S_2^i S_2^j )^{(2)} \,
( \hat x^i \hat x^j )^{(2)} \right\} \,,
\\[3pt]
\label{calH4D}
\tildeH^{(4)}_{\rm D} =& \;
\frac{2 \pi Z \alpha}{3} r_{E1}^2 \left \{ \delta^{(3)}(\vec x\,) - 
\lambda^2 \frac{\ee^{-\lambda r}}{4 \pi r} \right \} 
\nonumber\\
& \; + 
\frac{2 \pi Z \alpha}{3} r_{E2}^2 \left \{ \delta^{(3)}(\vec x\,) - 
\lambda^2 \frac{\ee^{-\lambda r}}{4 \pi r} \right \} \,.
\end{align}
\end{subequations}
These terms, in the limit $\lambda \rightarrow 0$,
combine to give the two-particle interaction
Hamiltonian of Eq.~\eqref{H4}. The spin-orbit
and magnetic terms in the limit $\tilde g_i \rightarrow 2$ and with $\lambda$
nonzero agree with the results given
in Eq.~(63) of \cite{AdJe2024gauge},
provided we reinterpret the spin operators 
appropriately, in a form applicable for 
spin-$1/2$ particles. 
Namely, for spin-$1/2$ point particles,
one needs to replace $r_{Ei}^2 \rightarrow
\frac{3}{4 m_i^2}$, as suggested in Eq.~(6) of \cite{ZaPaPa2022}.
In the general case, one uses 
the spin-dependent relations given in Eq.~\eqref{DFterm}
in order to relate the radius terms 
$r_{E1}^2$ and $r_{E2}^2$ with the mean-square charge
radii of the two particles.

%
%
\section{Applications}
\label{sec5}

%
%
\subsection{Spinless Bound Systems: Pionium}

One central point of our investigations 
is to derive general formulas for the 
relativistic-recoil correction to eVP,
applicable to bound Coulomb systems
with constituent spin-0, spin-$1/2$ and spin-$1$ particles.
One might think that this task has already been 
accomplished: Simply, one replaces, 
in Eq.~\eqref{calH4}, the mass terms $m_1$ and $m_2$
by the masses of the constituent particles,
and the spin operators accordingly,
and obtains the corresponding Hamiltonian.

However, there are a few subtleties 
involved in this process [in part, 
connected with Eq.~\eqref{DFterm}], and in the 
current section, we shall discuss these
by way of example, considering
particular bound systems of special interest.
First, we investigate pionium~\cite{KoRa2000,GaLyRuGa2001,JeSoIn2002,%
Sc2004plb,Sc2004epjc,Sc2005,GaLyRu2008,GaLyRu2009,DI2005,DI2015},
as an example of a bound system
consisting of two spinless particles
($\vec S_1 = \vec S_2 = \vec 0$).
For pionium, the two masses are equal
($m_1 = m_2 = m_\pi$), and 
there is no fine-structure, and no hyperfine
structure (bound states are characterized
by the principal quantum number $n$ and the 
orbital angular momentum $L$).
Furthermore, 
because of the spinless character of the pions,
we can replace $\tildeH^{(4)}_{\SO} \to 0$,
$\tildeH^{(4)}_{\FSS} \to 0$.
Furthermore, pions do not have a intrinsic
quadrupole moment, hence $Q_{E1} = Q_{E2} = 0$ and
$\tildeH^{(4)}_{\rm Q} \to 0$.
The magnetic interaction, in pionium, becomes
\begin{multline}
\label{HMpionium}
\tildeH^{(4)}_{\rm M} \to - \frac{Z \alpha}{2 m_\pi^2} 
p^i \frac{\ee^{-\lambda r}}{r} \Big [ \delta^{i j} + 
(1+\lambda r) \hat x^i \hat x^j  \Big ] p^j 
\\
- \frac{Z \alpha \, \lambda^2 }{4 m_\pi^2} 
\frac{\ee^{-\lambda r}}{r} 
\left ( 1 - \frac{\lambda r}{2} \right ) \,.
\end{multline}
In view of the absence of the Darwin--Foldy 
term for spinless particles, the Darwin term 
$\tildeH^{(4)}_{\rm D} \to \tildeH^{(4)}_{\rm FR}$
is reduced to the vacuum-polarization correction to the 
finite-radius (FR) effect,
\begin{equation}
\label{HDpionium}
\tildeH^{(4)}_{\rm FR} \to
\frac{4 \pi Z \alpha}{3} r_\pi^2 \left \{ \delta^{(3)}(\vec x\,) - 
\lambda^2 \frac{\ee^{-\lambda r}}{4 \pi r} \right \} \,.
\end{equation}
In the above formulas, we leave the nuclear charge
number $Z$ in the formulas, in order to 
facilitate the differentiation of the Coulomb-binding
parameter $Z\alpha$ from the radiative 
QED parameter $\alpha/\pi$, even 
if, for pionium, we have $Z=1$ (of course).

The term in Eq.~\eqref{HDpionium}
is the analog of the term studied in
Eq.~(65) of Ref.~\cite{Pa1996mu},
in the context of muonic hydrogen (there is
an additional prefactor $2$ in pionium
because because both constituent particles
have a finite radius).
Lattice calculations~\cite{GaEtAl2021}
yield a value $r_\pi = 0.648(15) \, {\rm fm}$
and we mention a recent experimental result
of $r_\pi = 0.640(7) \, {\rm fm}$ from 
from Ref.~\cite{CuBiRoSc2021}.
According to the latest
data compilation~\cite{PDG2024} of the particle
data group, the best current
estimates for the 
mass and the charge radius of the pion are
\begin{equation}
\label{param}
m_\pi = 139.57039(18) \, {\rm MeV} \,,
\qquad
r_\pi = 0.659(4) \, {\rm fm} \,.
\end{equation}
We shall use these values for the 
numerical results given in Table~\ref{table2}.
As with all corrections studied here,
we recall that it is necessary
to consider a first-order perturbation-theory
effect, given by the expectation value of the
operators in Eq.~\eqref{treeH4},
and a second-order perturbation-theory
effect, given by a vacuum-polarization
correction to the wave function,
described by the Schr\"{o}dinger--Coulomb Green function
and the corresponding Breit operator
from Eq.~\eqref{H4}. 

\begin{table}[t!]
\caption{\label{table2} The first-
and second-order recoil corrections 
to the eVP energy shift in pionium
are given for $1S$ and $2S$ states.
The total recoil
correction is obtained as
the sum of the first-order
and second-order terms: $E_{\eVP}^{(\rm RR)} = E^{(1)} + E^{(2)}$.
All entries are given in units of meV.}
\def\arraystretch{1.7}
\begin{tabular}{c@{\hspace*{0.6cm}} 
  S[table-format = 1.6e+1] @{\hspace*{0.3cm}} 
  S[table-format = 1.6e+1] }
\hline
\hline
& \multicolumn{1}{l}{$1S$} 
& \multicolumn{1}{l}{$2S$} \\
\hline
$E^{(1)}_{\rm M}$  &    -0.04815468(11)  & -0.006678205(16) \\
$E^{(1)}_{\rm FR}$ &     0.02427(30)    &   0.003123(38)    \\
$E^{(1)}$          &    -0.02389(30)    &  -0.003555(38)    \\
\hline
$E^{(2)}_{\rm K+M}$  &  -0.25407049(49)  & -0.030929877(74) \\
$E^{(2)}_{\rm FR}$ &     0.04495(55)    &   0.004720(57)    \\
$E^{(2)}$          &    -0.20912(55)    &  -0.026209(57)    \\
\hline
$E_{\eVP}^{(\rm RR)}$ & -0.23301(62)    &  -0.029764(69) \\
\hline
\hline
\end{tabular}
\end{table}

The first-order contributions to the recoil
correction to eVP are 
\begin{subequations}
\begin{align}
E^{(1)}_X(n L) =& \; \langle n L | H^{(5)}_{X} | n L \rangle \,,
\\
E^{(1)}(n L) =& \; \langle n L | H^{(5)}_{\eVP} | n L \rangle =
E^{(1)}_{\rm M} + E^{(1)}_{\rm FR}  \,.
\end{align}
\end{subequations}
where $X \in \{ {\rm M}, {\rm FR} \}$,
and the Hamiltonians are given in 
Eqs.~\eqref{trafo45},~\eqref{calH5},~\eqref{HMpionium}
and~\eqref{HDpionium}.

The nonretarded Breit Hamiltonian $H^{(4)}$, for pionium,
is the sum of a kinetic$+$magnetic term $H^{(4)}_{\rm K+M}$,
and a finite-radius term $H^{(4)}_{\rm FR}$,
\begin{subequations}
\begin{align}
H^{(4)} \to & \; H^{(4)}_{\rm K+M} + H^{(4)}_{\rm FR} \,,
\\
H^{(4)}_{\rm K+M} = & \;
- \frac{p^4}{4 m_\pi^3} 
- \frac{Z \alpha}{2 m_\pi^2} \, p^i \,
\left ( \frac{\delta^{i j} + \hat x^i \hat x^j}{r} \right ) p^j \, , 
\\
H^{(4)}_{\rm FR} = & \;
\frac{4 \pi Z \alpha}{3} \, r_\pi^2 \, \delta^{(3)}(\vec x\,) \,.
\end{align}
\end{subequations}
The spinless, nonretarded Breit Hamiltonian
can be obtained from Eqs.~\eqref{H4},~\eqref{H4suppl},
and~\eqref{treeH4} by setting to zero all 
spin operators, and setting $m_1 = m_2 = m_\pi$.
Its explicit form has been given in Eq.~(15) of 
Ref.~\cite{JeSoIn2002}.
Regarding the matrix elements of the spinless,
nonretarded Breit Hamiltonian, we refer
to Eq.~(38) of Ref.~\cite{AdJe2025prr},
Eq.~(16) of Ref.~\cite{JeSoIn2002},
Eq.~(5.13) of Ref.~\cite{Sc2004epjc}, and
Eqs.~(5.30) and (8.2) of Ref.~\cite{GaLyRu2008}.

With the definition of the one-loop eVP potential
(Uehling potential) as
\begin{equation}
\label{V1eVP}
V^{(1)}_{\eVP}(r) = 
\frac{\alpha}{\pi} \int_0^1 \dd v \, f_1(v) \,
\left( - \frac{\alpha}{r} \ee^{-\lambda r} \right) \,,
\end{equation}
where $\lambda \equiv \lambda(v) =
2 m_e/\sqrt{1-v^2}$ as defined in Eq.~\eqref{deflambda},
one obtains a second-order effect,
\begin{subequations}
\begin{align}
E^{(2)}_{\rm X}(n L) =& \; 2 \, \langle n L |
H^{(4)}_{\rm X} \, \left( \frac{1}{E_{\rm S} - H_{\rm S}} \right)' \,
V^{(1)}_{\eVP} | n L \rangle \,,
\\
E^{(2)}(n L) =& \; 2 \, \langle n L |
H^{(4)} \, \left( \frac{1}{E_{\rm S} - H_{\rm S}} \right)' \,
V^{(1)}_{\eVP} | n L \rangle 
\nonumber\\
=& \; E^{(2)}_{\rm K+M} + E^{(2)}_{\rm FR} \,,
\end{align}
\end{subequations}
where $X \in \{ {\rm K+M}, {\rm FR} \}$.
The unperturbed Schr\"{o}dinger energy is 
\begin{equation}
E_{\rm S} = -\frac{\alpha^2 m_\pi}{4 n^2} \,,
\end{equation}
and the Schr\"{o}dinger--Coulomb Hamiltonian is
\begin{equation}
H_{\rm S} =  \frac{\vec p^{\,2}}{m_\pi} - \frac{\alpha}{r} \, .
\end{equation}
Results for the first-order
and second-order contributions
are given in Table~\ref{table2},
where the theoretical uncertainties
correspond to the numerical values of the 
charged pion's mass and charge radius from Eq.~\eqref{param}.

%
%
\subsection{Fine--Structure of Muonic Hydrogen}
\label{appb}

For muonic hydrogen, one encounters the 
masses
\begin{equation}
m_1 = m_\mu \,, \qquad m_2 = m_N = m_p \,.
\end{equation}
Here, $m_\mu$ is the muon mass, and $m_N = m_p$ 
is the nuclear mass.
We attempt to reconcile the radiatively 
corrected Breit Hamiltonian, given in Eq.~\eqref{calH4},
with the Hamiltonian used in 
Refs.~\cite{Pa1996mu,VePa2004,Je2011pra,KaIvKo2012,AdJe2024gauge},
for the evaluation of relativistic 
and recoil corrections to vacuum polarization
in muonic hydrogenlike ions.
For the discussion of the 
fine-structure, we can exclude hyperfine
effects (due to the nuclear spin).

There is slight subtlety:
even for a point-like proton, 
the Darwin term in Eq.~\eqref{calH4D} 
gives a nonvanishing contribution
originating from the proton,
because of the definition
of the charge radius on which 
Eqs.~\eqref{treeH4D} and~\eqref{calH4D} are 
based. Namely, for a pointlike nucleus,
one has
$r_{E2}^2 \to \frac{3 \delta_I }{4 m_N^2}$,
according to Eq.~\eqref{DFterm}.
The following replacements are
thus relevant for the muonic systems,
\begin{equation}
\label{rp}
r_{E1}^2 \to \frac{3}{4 m_\mu^2} \,,
\quad
r_{E2}^2 \to \frac{3}{4 m_p^2} + r_p^2 \,,
\quad
\tilde g_1 \to 2 \,.
\end{equation}
where $r_p^2$ is the mean-square proton charge radius.
One then takes into account the fact that the 
orbiting particle is a spin-$1/2$ particle,
ignores hyperfine effects (one replaces $\vec S_2 \to \vec 0$),
and observes that the intrinsic quadrupole 
moments of the muon and proton vanish,
\begin{equation}
\vec S_1 \to \frac{\vec\sigma_\mu}{2} \,,
\quad 
Q_{E1} \to 0 \,,
\quad 
Q_{E2} \to 0 \,.
\end{equation}
Here $\vec \sigma_\mu$ is the 
vector of Pauli spin matrices for the muon spin.
Hence, for the fine-structure of 
muonic hydrogen, we can replace
$\tildeH^{(4)}_{\rm Q} \to 0$,
and we can also replace 
the Fermi spin-spin interaction 
by $\tildeH^{(4)}_{\FSS} \to 0$.        

For muonic hydrogen,
the magnetic term takes the form
\begin{multline}
\tildeH^{(4)}_{\rm M} \to - \frac{Z \alpha}{2 m_\mu m_p}
p^i \frac{\ee^{-\lambda r}}{r} \Big [ \delta^{i j} +
(1+\lambda r) \hat x^i \hat x^j  \Big ] p^j
\\
- \frac{Z \alpha \lambda^2 }{4 m_\mu m_p}
\frac{\ee^{-\lambda r}}{r}
\left ( 1 - \frac{\lambda r}{2} \right ) \,.
\end{multline}
The latter form is equal to the sum 
$\delta \tildeH_2 + \delta \tildeH_3$,
used in their respective forms
defined in Eqs.~(63d) and (63e) 
of Ref.~\cite{AdJe2024gauge}.
The spin-orbit term takes the form
\begin{multline}
\tildeH^{(4)}_{\SO} \to Z \alpha \,
\left( \frac{1}{4 m_\mu^2} + 
\frac{1}{2 m_\mu m_p} \right) \,
\frac{\ee^{-\lambda r} }{r^3} 
\\
\times 
(1+\lambda r) \,
\vec \sigma_\mu \cdot \vec L 
\equiv 
\delta \tildeH_4 \,,
\end{multline}
where $\delta \tildeH_4$ is used in the notation
of Eq.~(63f) of Ref.~\cite{AdJe2024gauge}.

Now we investigate the Darwin term.
One verifies that, for muonic hydrogen-like ions,
the individual terms listed in Eq.~\eqref{calH4D}
can be expressed as a sum of two terms,
\begin{equation}
\tildeH^{(4)}_D =
\tildeH^{(4)}_{D,1} +
\tildeH^{(4)}_{D,2} \,.
\end{equation}
The first term is
\begin{multline}
\tildeH^{(4)}_{\rm D,1} =
\frac{\pi Z\alpha}{2} \left( \frac{1}{m_\mu^2} + 
\frac{1}{m_p^2} \right) 
\\
\times \left \{ \delta^{(3)}(\vec r\,) -
\lambda^2 \frac{\ee^{-\lambda r}}{4 \pi r} \right \} 
\equiv \delta \tildeH_1 \,,
\end{multline}
where $\delta \tildeH_1$ is used in the notation
of Eq.~(63c) of Ref.~\cite{AdJe2024gauge}.
The finite-size contribution is
\begin{multline}
\tildeH^{(4)}_{\rm D,2} =
\frac{2 \pi Z\alpha}{3} r_p^2 
\left \{ \delta^{(3)}(\vec r\,) -
\lambda^2 \frac{\ee^{-\lambda r}}{4 \pi r} \right \}
\equiv \delta \tildeH_{\rm FS} \,,
\end{multline}
as discussed in Eq.~(65) of Ref.~\cite{Pa1996mu}.
We have checked our results above
against corresponding entries in Ref.~\cite{AdJe2024gauge},
and references therein. 
For the comparison to the literature, 
it is important to realize that,
{\em e.g.}, the detailed breakdown of the 
corrections to the fine-structure of muonic hydrogen
in Ref.~\cite{Je2011pra} excludes the 
finite-radius correction, which is 
treated, in both first- and second-order 
perturbation theory, 
in Eqs.~(65)~and~(66) of Ref.~\cite{Pa1996mu}.

The history of the recoil correction to eVP involves
gradual progress in the calculations. For example, in
Ref.~\cite{KaIvKo2012}, it was suggested that in 
Ref.~\cite{Bo2011preprintv5}, some gauge-dependent terms 
were present in the Hamiltonian which, along with a numerical error, prevented 
agreement with \cite{Je2011pra} for the $n=2$ 
Lamb shift in muonic hydrogen. This fact has recently
been clarified in Ref.~\cite{PaPa2025}, in the text
preceding Eq. (21) of that reference, pointing out that 
a corresponding statement in \cite{AdJe2024gauge} should properly refer only
to the result of \cite{Bo2011preprintv5} and not to \cite{VePa2004}.


%
%
\subsection{Hyperfine Structure of Muonic Hydrogen}

For the hyperfine structure of muonic hydrogen,
we now investigate terms which involve the 
spin operator of the proton, 
$\vec S_2 = \vec S_p = \vec\sigma_p/2$.
The $g$ factor of the bound muon
is $\tilde g_1 = \tilde g_\mu = 2$.
Because the nuclear magneton is defined 
with respect to the proton mass, we have
\begin{equation}
\tilde g_2 \to \tilde g_p = g_p 
= 5.585\,694\,6893(16) \,,
\end{equation}
where we use the value from Ref.~\cite{MoNeTaTi2025}.
We recall the spin operator of the muon as
$\vec S_\mu = \vec\sigma_\mu/2$.
Neither the muon nor the proton have intrinsic 
quadrupole moments, hence, for the 
hyperfine structure of muonic hydrogen,
we can replace $\tildeH^{(4)}_{\rm Q} \to 0$.

Terms that involve $\vec \sigma_p$ are 
the spin-orbit term
\begin{multline}
\tildeH^{(4)}_{\SO} \to
\frac{Z \alpha \,
\ee^{-\lambda r} }{4 r^3} (1+\lambda r) 
\left( \frac{g_p-1}{m_p^2}  + 
\frac{g_p}{m_\mu m_p} \right) 
\vec \sigma_p \cdot \vec L \,, 
\end{multline}
and the Fermi spin-spin term
\begin{multline}
\tildeH^{(4)}_{\FSS} \to \frac{\pi  Z \alpha \, g_p}{3 m_\mu m_p}
\vec \sigma_\mu \cdot \vec \sigma_p \,
\left\{ \delta^{(3)}(\vec x \, )
- \lambda^2 \frac{\ee^{-\lambda r}}{4 \pi r} \right\}
\\
+ \frac{3  Z \alpha \, g_p}{8 m_\mu m_p} \,
\frac{\ee^{-\lambda r}}{r^3}
\left(1 + \lambda r + \frac{\lambda^2 r^2}{3} \right) \,
\\
\times
( \sigma_\mu^i \sigma_p^j )^{(2)}  \; ( \hat x^i \hat x^j )^{(2)} \,.
\end{multline}
For $P$ states and states with higher 
orbital angular momenta, we recall that
the fine-structure and hyperfine-structure
Hamiltonians do not commute, which 
leads to mixing terms [see Eq.~(91)
of Ref.~\cite{Pa1996mu}].
We have checked our formulas,
and corresponding results for the hyperfine
structure of muonic hydrogen,
for $1S$ and $2S$ states, against the 
literature [Eqs.~(18) and (30) of Ref.~\cite{Ma2005mup}].

%
%
\subsection{Fine--Structure of Muonic Deuterium}

For muonic deuterium, one encounters the 
masses
\begin{equation}
m_1 = m_\mu \,, \qquad m_2 = m_d \,.
\end{equation}
Here, $m_\mu$ is the muon mass, and $m_d$ 
is the deuteron mass.
We first ignore the nuclear 
spin and concentrate on the fine-structure.
For muonic deuterium, with its
spin-1 nucleus, we can replace
[see also the discussion surrounding 
Eq.~\eqref{rp}]
\begin{equation}
r_{E1}^2 \to \frac{3}{4 m_\mu^2} \,,
\quad
r_{E2}^2 \to r_d^2 \,,
\quad
\tilde g_1 \to 2 \,.
\end{equation}
One then takes into account the fact that the 
orbiting particle is a spin-$1/2$ particle, so
\begin{equation}
\vec S_1 \to \frac{\vec\sigma_\mu}{2} \,,
\end{equation}
ignores hyperfine effects,
and observes that the intrinsic quadrupole 
moment of the muon vanishes
while that of the deuteron is nonzero.
Since we are considering the fine-structure,
we can ignore terms involving the deuteron
spin ($\vec S_2 = \vec S_d$).
For the deuteron,
we use a root-mean-square deuteron charge radius~\cite{MoNeTaTi2025}
of
\begin{equation}
\label{rd}
r_d = 2.12778(27) \times 10^{-15} \, {\rm fm} \,.
\end{equation}
For the deuteron's quadrupole moment (divided by the
elementary charge), we use the value 
of~\cite{KeRaRaZa1939,KeRaRaZa1940,PuKoPa2020})
\begin{equation}
\label{QEd}
Q_{Ed} = 0.285\,699(24) \, {\rm fm}^2 \,.
\end{equation}
For the fine-structure,
we can replace $\vec S_2 \to \vec 0$, and
we can ignore the quadrupole terms in muonic hydrogenlike ions even if the nucleus 
has an intrinsic quadrupole moment ($\tildeH^{(4)}_{\rm Q} \to 0$).
The Fermi spin-spin interaction 
can also be ignored, $\tildeH^{(4)}_{\FSS} \to 0$.
One verifies that, for muonic hydrogen-like ions,
the individual terms listed in Eq.~\eqref{calH4M}
and Eq.~\eqref{calH4SO} take the form
\begin{multline}
\tildeH^{(4)}_{\rm M} \to - \frac{Z \alpha}{2 m_\mu m_d} 
p^i \frac{\ee^{-\lambda r}}{r} \Big [ \delta^{i j} + 
(1+\lambda r) \hat x^i \hat x^j  \Big ] p^j 
\\
- \frac{Z \alpha \, \lambda^2 }{4 m_\mu m_d} 
\frac{\ee^{-\lambda r}}{r} 
\left ( 1 - \frac{\lambda r}{2} \right ) \,, 
\end{multline}
\begin{multline}
\tildeH^{(4)}_{\SO} \to \frac{Z \alpha \,
\ee^{-\lambda r} }{4 r^3} (1+\lambda r) 
\left\{ \left( \frac{1}{m_\mu^2}  + 
\frac{ 2 }{m_\mu m_d} \right) 
\vec \sigma_\mu \cdot \vec L \right\} \,.
\end{multline}
The Darwin term listed in Eq.~\eqref{calH4D}
can be written as the sum of two individual 
contributions,
\begin{equation}
\tildeH^{(4)}_D =
\tildeH^{(4)}_{D,1} +
\tildeH^{(4)}_{D,2} \,.
\end{equation}
The first of these is a generalized form of the 
{\em zitterbewegung} term of the muon,
which is obtained here as the Darwin--Foldy (DF)
term contributing to the finite-radius 
term for spin-$1/2$ particles~\cite{PaKa1995,Je2011radii},
\begin{equation}
\tildeH^{(4)}_{\rm D,1} = 
\tildeH^{(4)}_{\rm DF} = 
\frac{\pi Z \alpha}{2 m_\mu^2} \left \{ \delta^{(3)}(\vec x\,) - 
\lambda^2 \frac{\ee^{-\lambda r}}{4 \pi r} \right \} \,,
\end{equation}
while the second one is a vacuum-polarization
correction to the finite-size effect,
which, in the current context, we refer to 
as the finite-radius (FR) effect,
\begin{equation}
\tildeH^{(4)}_{\rm D,2} = 
\tildeH^{(4)}_{\rm FR} = 
\frac{2 \pi Z \alpha}{3} r_{d}^2 \left \{ \delta^{(3)}(\vec x\,) - 
\lambda^2 \frac{\ee^{-\lambda r}}{4 \pi r} \right \} \,.
\end{equation}
We have checked the eVP-recoil correction
obtained from the above formulas 
(for muonic deuterium) against the corresponding
entries in Table~I of Ref.~\cite{AdJe2024gauge},
and in Table~IV of Ref.~\cite{KaIvKo2012},
for all states with principal quantum numbers
$1 \leq n \leq 4$.
One should remember, though, that the 
entries in Table~I of Ref.~\cite{AdJe2024gauge},
and in Table~IV of Ref.~\cite{KaIvKo2012},
{\em exclude} the finite-radius correction
due to $\tildeH^{(4)}_{\rm FR}$,
and also the finite-radius term
in the nonretarded Breit Hamiltonian, {\em i.e.},
the term proportional to the charge 
radius $r_{E2}^2 = r_d^2$ in Eq.~\eqref{DFterm}.
We take the opportunity to point out 
that the entry in Eq.~(43) of Ref.~\cite{KrMa2011}
for the second-order contribution to the 
$2P_{1/2}$--$2S_{1/2}$ Lamb shift in muonic deuterium appears to 
be in need of a correction,
replacing $0.0530$\,meV $\to$ $0.056984$\,meV.
After the replacement, the sum of the 
results given in Eqs.~(27), (28), (29), (30)
and (43) of Ref.~\cite{KrMa2011}
is in agreement with the difference 
of the corresponding entries in 
Table~I of Ref.~\cite{AdJe2024gauge},
$[-6.3675 \times 10^{-3} - ( - 2.8149 \times 10^{-2} )]$\,meV =
$0.021781$\,meV, and also with the
entry in Table~IV of Ref.~\cite{KaIvKo2012}.

%
%
\subsection{Hyperfine Structure of Muonic Deuterium}

We now consider those terms in 
Eq.~\eqref{calH4} which involve the 
spin operator of the deuteron,
$\vec S_2 = \vec S_d$.
The magnetic term and the 
Darwin term are independent of the 
nuclear spin, hence, for the 
consideration of the hyperfine structure,
we can replace $\tildeH^{(4)}_{\rm M} \to 0$,
and $\tildeH^{(4)}_{\rm D} \to 0$.
The spin-orbit term becomes
\begin{multline}
\tildeH^{(4)}_{\SO} \to \frac{Z \alpha \,
\ee^{-\lambda r} }{2 r^3} (1+\lambda r) 
\\
\times \left\{ 
\left( \frac{\tilde g_d-1}{m_d^2}  + 
\frac{\tilde g_d}{m_\mu m_d} \right) 
\vec S_d \cdot \vec L  \right\} \,, 
\end{multline}
while the Fermi spin-spin term becomes
\begin{multline}
\tildeH^{(4)}_{\FSS} \to \frac{2 \pi  Z \alpha \,  \tilde g_d}{3 m_\mu m_d}
\vec \sigma_\mu \cdot \vec S_d \,
\left\{ \delta^{(3)}(\vec x \, )
- \lambda^2 \frac{\ee^{-\lambda r}}{4 \pi r} \right\}
\\
+ \frac{3  Z \alpha \,  \tilde g_d}{4 m_\mu m_d} \,
\frac{\ee^{-\lambda r}}{r^3}
\left(1 + \lambda r + \frac{\lambda^2 r^2}{3} \right) 
\\
\times ( \sigma_\mu^i S_d^j )^{(2)}  \; ( \hat x^i \hat x^j )^{(2)} \, ,
\end{multline}
and the quadrupole term becomes
\begin{multline}
\tildeH^{(4)}_{\rm Q} \to - \frac{3 Z \alpha}{2} 
\frac{\ee^{-\lambda r}}{r^3} 
\left ( 1+\lambda r + \frac{\lambda^2 r^2}{3} \right ) 
\\
\times
Q_{Ed} \, ( S_d^i S_d^j )^{(2)} \,
( \hat x^i \hat x^j )^{(2)} \,.
\end{multline}
Fine-structure and hyperfine-structure
effects, as well as nuclear-polarization 
corrections in muonic deuterium,
have received considerable attention
recently~\cite{Pa2011,KaPaYe2018,KrMa2011,FaMaMaSo2014,FaMaMaSo2015,JiZhPl2024}.
The mixing matrix of the 
fine-structure and hyperfine-structure
components has been discussed in
Eq.~(45) of Ref.~\cite{KrEtAl2016deuterium}
and in Eq.~(62) of Ref.~\cite{FaMaMaSo2015}.
We leave the nuclear charge number $Z$ in the formulas,
so that the above formulas can easily 
be generalized to muonic ions with 
a spin-$1$ nucleus.
We have checked our formulas above, for 
the hyperfine structure of muonic deuterium,
against results presented in Eqs.~(11),~(12), 
(22) and (23) of Ref.~\cite{FaMaMaSo2014}.

%
%
\section{Numerical Values for Deuteronium}
\label{sec6}

Let us now focus on the bound system of a 
deuteron and its antiparticle, which we refer
to as deuteronium~\cite{AdJe2025prl,AdJe2025prr}.
The deuteron mass is denoted as $m_d$,
and the reduced mass of the system is $m_r = m_d/2$.
Values for the deuteron radius
and its quadrupole moment have been
given in Eqs.~\eqref{rd} and~\eqref{QEd}.
Appropriate quantum numbers in deuteronium are 
the principal quantum number $n$, the total orbital
angular momentum $L$, the total spin $S$, 
and the total angular momentum quantum number $J$
(although there is mixing of states with $L=J$, as discussed below.)
We use spectroscopic notation $n {}^{2S + 1} L_J$,
where $L$ is denoted by the letters $S$ (for $L = 0$),
$P$ (for $L = 1$), $D$ (for $L=2$), and so on.
As an example, for the state with quantum numbers
$n = 3$, $L = 2$, $S = 1$, $J = 2$,
the spectroscopic notation is $3 {}^3\! D_2$.

Let us break down the individual contributions
and start with the first-order perturbation theory
terms, which are given by the expectation 
value of the radiatively corrected 
Breit Hamiltonian $H^{(5)}_{\eVP}$ 
given in Eqs.~\eqref{trafo45} and~\eqref{calH5}.  
We define the individual first-order perturbative 
terms [denoted with the superscript $(1)$] as follows,
\begin{multline}
\label{E1X}
E^{(1)}_{\rm X}(n {}^{2S+1}L_J) = 
\frac{\alpha}{\pi} \int_0^1 \dd v \, f_1(v) \, 
\\
\times \langle n {}^{2S+1}L_J | \tildeH^{(4)}_{\rm X} | n {}^{2S+1}L_J \rangle \,,
\end{multline}
%
where we consider the terms with
\begin{equation}
{\rm X} \in \{ \Mone, \Mtwo, \SO, \FSSone,\FSStwo, {\rm Q}, {\rm D} \} 
\end{equation}
individually.
For deuteronium, we have~\cite{AdJe2025prl,AdJe2025prr}
\begin{equation}
m_1 = m_2 = m_d \,, \qquad
Z = 1 \,,
\end{equation}
where $m_d$ is the deuteron mass.
We split the magnetic term into two contributions,
$\tildeH^{(4)}_{\rm M} = \tildeH^{(4)}_{\Mone} +
\tildeH^{(4)}_{\Mtwo}$, where
\begin{subequations}
\begin{align}
\tildeH^{(4)}_{\Mone} =& \; - \frac{\alpha}{2 m_d^2}
p^i \frac{\ee^{-\lambda r}}{r} \Big [ \delta^{i j} +
(1+\lambda r) \hat x^i \hat x^j  \Big ] p^j,
\\
\tildeH^{(4)}_{\Mtwo} =& \; 
- \frac{\alpha \lambda^2 }{4  m_d^2}
\frac{\ee^{-\lambda r}}{r}
\left ( 1 - \frac{\lambda r}{2} \right ) \,   .
\end{align}
\end{subequations}
The spin-orbit contribution is
\begin{equation}
\tildeH^{(4)}_\SO = \frac{\alpha}{2 m_d^2} (2 \tilde g_d-1) \frac{e^{-\lambda r}}{r^3} (1+\lambda r) \, \vec S \cdot \vec L \, ,
\end{equation}
where $\vec S = \vec S_1 + \vec S_2$ is the total spin operator.
We consider the two parts of the Fermi spin-spin term separately,
$\tildeH^{(4)}_{\FSS} = \tildeH^{(4)}_{\FSSone} +\tildeH^{(4)}_{\FSStwo}$, where
\begin{subequations}
\begin{align}
\tildeH^{(4)}_{\FSSone} =&  \frac{2 \pi \alpha \tilde g_d^2}{3 m_d^2}
\left\{ \delta^{3}(\vec r \, ) - \lambda^2 \frac{\ee^{-\lambda r}}{4 \pi r} \right\}
\vec S_1 \cdot \vec S_2 \, ,
\\
\tildeH^{(4)}_{\FSStwo} =&  \frac{3 \alpha \tilde g_d^2}{4 m_d^2}
\frac{\ee^{-\lambda r}}{r^3}
\left ( 1 + \lambda r + \frac{\lambda^2 r^2}{3} \right )
( S_1^i S_2^j )^{(2)}   ( \hat x^i \hat x^j )^{(2)}   .
\end{align}
\end{subequations}
The quadrupole term is
\begin{eqnarray}
\tildeH^{(4)}_{\rm Q} &=& - \frac{3 \alpha}{2 m_d^2} \tilde Q_d
\frac{\ee^{-\lambda r}}{r^3}
\left ( 1 + \lambda r + \frac{\lambda^2 r^2}{3} \right ) \nonumber \\
&\times&
\left [ ( S_1^i S_1^j )^{(2)} + ( S_2^i S_2^j )^{(2)} \right ]  ( \hat x^i \hat x^j )^{(2)}  \, ,
\end{eqnarray}
where the dimensionless quadrupole moment factor $\tilde Q_d$ is defined as
\begin{equation}
\tilde Q_d \equiv m_d^2 Q_d = \left ( \frac{m_d c^2}{\hbar c} \right )^2 Q_d = 25.8120(22) \, .
\end{equation}
Finally, the Darwin (finite size) contribution is
\begin{equation}
\tildeH^{(4)}_{\rm D} = \frac{4 \pi \alpha}{3 m_d^2} \, \tilde r_d^2 \, 
\left \{ \delta^{(3)}(\vec r) - \lambda^2 \frac{\ee^{-\lambda r}}{4 \pi r} \right \} \, ,
\end{equation}
where the dimensionless radius is
\begin{equation}
\tilde r_d \equiv m_d r_d = \frac{m_d c^2}{\hbar c} r_d = 20.2248(26) \, .
\end{equation}

The matrix elements are to be taken with 
the states
$\langle \vec r \, | n L S J M_J \rangle =
\psi_{n L S J M_J}(\vec r \, )$, 
where the wave functions are
\begin{equation}
\psi_{n L S J M_J}(\vec r \, ) = R_{nL}(r) \,
\Xi^{L S}_{J M_J}(\theta, \varphi) \,,
\end{equation}
and where $R_{nL}(r)$ is the nonrelativistic 
Schr\"{o}dinger--Coulomb wave function 
for the appropriate reduced mass $m_r = m_d/2$
and nuclear charge number $Z = 1$
(see Chap.~4 of Ref.~\cite{JeAd2022book}).
Furthermore, the spin-angular
function $\Xi^{L S}_{J M_J}(\theta, \varphi)$
can be expressed as
\begin{multline}
\Xi^{L S}_{J M_J}(\theta, \varphi) =
\sum_{M_L M_S m_a m_b}
C^{J M_J}_{L M_L S M_S} \;
C^{S M_S}_{1 m_a 1 m_b} 
\\
\times Y_{L M_L}(\theta, \varphi) \; \chi^{(1)}_{m_a} \; \chi^{(2)}_{m_b}
\end{multline}
where the Clebsch-Gordan coefficients
$C^{J M_J}_{j_1 m_1 j_2 m_2}$ are used
in the notation of Chap.~6 of Ref.~\cite{JeAd2022book},
and $\chi^{(1)}_{m_a}$ is the fundamental spin state
spinor for particle~$1$ (deuteron) 
with magnetic projection $m_a$,
and $\chi^{(2)}_{m_b}$ is the fundamental spin state
for particle~$2$ (antideuteron)
with magnetic projection $m_b$.
Expressions for the matrix elements of the various angular
operators appearing in the terms of $\tildeH^{(4)}$ were given in Appendix B of \cite{AdJe2025prr}.
We observe that the 
results for the first-order matrix elements,
given by $E^{(1)}_{\rm X}$ as outlined in Eq.~\eqref{E1X},
are independent of the projection $M_J$.

Deuteronium states with $S=0$, $L=J$ and $S=2$, $L=J$ are mixed at the level of
the Breit corrections (that is, at order $m_d \alpha^4$), as discussed in
\cite{AdJe2025prr}.  The mixed states and corresponding Breit energy
contributions were found by diagonalizing the mixing matrix

\begin{equation}
\langle n L S' J M_J \vert H^{(4)} \vert n L S J M_J \rangle = 
m_d \alpha^4 M^{n L J}_{S' S} \, ,
\end{equation}
as given explicitly in (37) of \cite{AdJe2025prr}.  (These matrix elements are
independent of $M_J$.). The only off-diagonal matrix elements have $S=0$, $L=J$
and $S=2$, $L=J$.  So, for each value of $L=J$, there are two mixed states.
The state with the smaller (larger) eigenvalue is labeled $n^-\!J_J$ ($n^+\!
J_J$).  For the order $m_d \alpha^5$ relativistic recoil corrections, we have
\begin{eqnarray}
E_\pm^{(5)} &=& \xi^*_\pm H^{(5)} \xi_\pm \nonumber \\
&\phantom{=}& \hspace{-1.8cm} = \sum_{\substack{S'=0,2\\S=0,2}} 
\xi^*_\pm(S') \langle n J S' J M_J \vert H^{(5)} \vert n J S J M_J \rangle \xi_\pm(S) \, ,
\end{eqnarray}
where the eigenvectors of the mixing matrix $\xi_\pm$ satisfy
\begin{equation}
m_d \alpha^4 \sum_{S=0,2} M^{n J J}_{S' S} \, \xi_\pm(S) = E_\pm^{(4)} \xi_\pm(S') \, .
\end{equation}
The Breit energies $E^{(4)}$ are given in Table III of \cite{AdJe2025prr}.

The $3D$ and $4D$ states of deuteronium have been identified as particularly 
interesting for spectroscopic investigations on New Physics effects~\cite{AdJe2025prl}.
We here selected the states  $3 {}^3 D_2$ and  $4 {}^3 D_2$ with spin $S=1$ and total angular momentum $J = 2$
for our illustrative calculations of the individual contributions.
For the state of deuteronium with the principal quantum number $n=3$,
and the spectroscopic notation $3 {}^3 D_2$, the individual results are
\begin{subequations}
\begin{align}
E^{(1)}_{\Mone}(3 {}^3 D_2) =& \; -0.003\,537 \, \meV \,,
\\
E^{(1)}_{\Mtwo}(3 {}^3 D_2) =& \; \phantom{xx}\, 0.000\,105 \, \meV \,,
\\
E^{(1)}_\SO(3 {}^3 D_2) =& \; -0.002\,201 \, \meV \,,
\\
E^{(1)}_{\FSSone}(3 {}^3 D_2) =& \; \phantom{xx}\, 0.000\,900 \, \meV \,,
\\
E^{(1)}_{\FSStwo}(3 {}^3 D_2) =& \, \phantom{xx}\, 0.001\,782 \, \meV \,,
\\
E^{(1)}_{\rm Q}(3 {}^3 D_2) =& \; \phantom{xx}\, 0.031\,309(3) \, \meV \,,
\\
E^{(1)}_{\rm D}(3 {}^3 D_2) =& \; -0.250\,691(64) \, \meV \,,
\\
E^{(1)}(3 {}^3 D_2) =& \; -0.222\,33(7) \, \meV \,.
\end{align}
\end{subequations}
For the state with $n = 4$, and $S=1$, $L =J = 2$,
with spectroscopic notation $4 {}^3 D_2$,
the individual contributions are
\begin{subequations}
\begin{align}
E^{(1)}_{\Mone}(4 {}^3 D_2) =& \; -0.001\,954 \, \meV \,,
\\
E^{(1)}_{\Mtwo}(4 {}^3 D_2) =& \; \phantom{xx}\, 0.000\,038 \, \meV \,,
\\
E^{(1)}_\SO(4 {}^3 D_2) =& \; -0.001\,043 \, \meV \,,
\\
E^{(1)}_{\FSS1}(4 {}^3 D_2) =& \; \phantom{xx}\, 0.000\,391 \, \meV \,,
\\
E^{(1)}_{\FSS2}(4 {}^3 D_2) =& \; \phantom{xx}\, 0.000\,826 \, \meV \,,
\\
E^{(1)}_{\rm Q}(4 {}^3 D_2) =& \; \phantom{xx}\, 0.014\,517\,1(12) \, \meV \,,
\\
E^{(1)}_{\rm D}(4 {}^3 D_2) =& \, -0.108\,783(28) \, \meV \,,
\\
E^{(1)}(4 {}^3 D_2) =& \; -0.096\,01(3) \, \meV \,.
\end{align}
\end{subequations}
The total first-order contribution is
\begin{equation}
\label{E1_formula}
E^{(1)}(n {}^{2S+1}L_J) =
\langle n {}^{2S+1}L_J | H^{(5)}_{\eVP} | n {}^{2S+1}L_J \rangle \,,
\end{equation}
where $H^{(5)}_{\eVP} $ is given in Eq.~\eqref{calH5}.
Results for the total first-order terms for 
$3P$, $3D$, $4D$ and $4F$ states
of deuteronium are given in Table~\ref{table3}.

\begin{table}[t!]
\caption{\label{table3} Expectation values 
$E^{(1)}$ of the radiatively corrected Breit 
Hamiltonian from Eq.~\eqref{calH5} 
(in meV) for $3P$, $3D$, $4D$ and $4F$ states of deuteronium.  
The dominant uncertainty affecting the $E^{(1)}$ energy contributions
is due to the uncertainty in the radius $r_d$.}
\def\arraystretch{1.2}
\begin{tabular}{c@{\hspace*{0.3cm}} 
  S[table-format = 1.6e+2] @{\hspace*{0.3cm}} 
  c @{\hspace*{0.3cm}}
  S[table-format = 1.6e+2] }
\hline
\hline
$3{}^{2S+1}P_J$ & 
\multicolumn{1}{l}{$E^{(1)}$} & 
$3{}^{2S+1}D_J$ & 
\multicolumn{1}{l}{$E^{(1)}$} \\
\hline
$3^{-}\!P_1$  &  -0.51655(5)   & $3^{- }\!D_2$ & -0.062271(5) \\
$3^{+}\!P_1$ &  -2.3859(8)     & $3^{+}\!D_2$ & -0.46433(13) \\
$3^{3}P_0$   &  -2.4681(4)     & $3^{3}D_1$   & -0.29292(7) \\
$3^{3}P_1$   &  -1.2899(4)     & $3^{3}D_2$   & -0.22233(7) \\
$3^{3}P_2$   &  -1.6910(4)     & $3^{3}D_3$   & -0.25828(7) \\
$3^{5}P_2$   &  -2.1630(4)     & $3^{5}D_0$   & -0.20918(7) \\
$3^{5}P_3$   &  -1.4601(4)     & $3^{5}D_1$   & -0.23650(7) \\
                     &                        & $3^{5}D_3$   & -0.28877(7) \\
                     &                        & $3^{5}D_4$   & -0.22935(7) \\
\hline
\hline
$4{}^{2S+1}D_J$ & 
\multicolumn{1}{l}{$E^{(1)}$} & 
$4{}^{2S+1}F_J$ & 
\multicolumn{1}{l}{$E^{(1)}$} \\
\hline
$4^{-}\!D_2$  &  -0.028705(2)   & $4^{-}\!F_3$  & -0.0049864(4) \\
$4^{+}\!D_2$ &  -0.20130(6)     & $4^{+}\!F_3$  & -0.042495(11) \\
$4^{3}D_1$ &  -0.12878(3)     & $4^{3}F_2$  & -0.025005(6) \\
$4^{3}D_2$ &  -0.09601(3)     & $4^{3}F_3$  & -0.020829(6) \\
$4^{3}D_3$ &  -0.11261(3)     & $4^{3}F_4$  & -0.023167(6) \\
$4^{5}D_0$ &  -0.08996(3)     & $4^{5}F_1$  & -0.020935(6) \\
$4^{5}D_1$ &  -0.10261(3)     & $4^{5}F_2$  & -0.022960(6) \\
$4^{5}D_3$ &  -0.12674(3)     & $4^{5}F_4$  & -0.024756(6) \\
$4^{5}D_4$ &  -0.09910(3)     & $4^{5}F_5$  & -0.021000(6) \\
\hline
\hline
\end{tabular}
\end{table}

Now, we investigate the second-order terms.
We define the individual contribution as follows,
\begin{multline}
\label{E2_formula}
E^{(2)}_{\rm X}(n {}^{2S+1}L_J) \\
= 2 \, \langle n {}^{2S+1}L_J | 
H^{(4)}_{\rm X} \left( \frac{1}{E_{\rm S} - H_{\rm S}} \right)' 
V^{(1)}_{\eVP} | n {}^{2S+1}L_J \rangle \,,
\end{multline}
where $\lambda \equiv \lambda(v) =
2 m_e/\sqrt{1-v^2}$ as defined in Eq.~\eqref{deflambda}.
Here
\begin{equation}
E_{\rm S} = -\frac{\alpha^2 m_d}{4 n^2}
\end{equation}
is the unperturbed Schr\"{o}dinger--Coulomb
energy and the Schr\"{o}dinger--Coulomb Hamiltonian is 
\begin{equation}
H_{\rm S} =  \frac{\vec p^{\,2}}{m_d} - \frac{\alpha}{r} \, .
\end{equation}
The individual terms $H_{\rm X}$ with 
\begin{equation}
{\rm X} \in \{ {\rm K}, {\rm M}, \SO, \FSS, {\rm Q} \} \,,
\end{equation}
in the Breit Hamiltonian are given in Eq.~\eqref{treeH4}.
Again, we devote special attention to the 
$3 {}^3 D_2$ and $4 {}^3 D_2$ states.
The individual results for $3 {}^3 D_2$ are  
\begin{subequations}
\begin{align}
E^{(2)}_{\rm K}(3 {}^3 D_2) =& \, -0.005\,359\,8 \, \meV \,,
\\
E^{(2)}_{\rm M}(3 {}^3 D_2) =& \; -0.013\,235\,1 \, \meV \,,
\\
E^{(2)}_\SO(3 {}^3 D_2) =& \; -0.004\,304\,1 \, \meV \,,
\\
E^{(2)}_{\FSS}(3 {}^3 D_2) =& \; \phantom{xx} 0.002\,603\,9 \, \meV \,,
\\
E^{(2)}_{\rm Q}(3 {}^3 D_2) =& \; \phantom{xx} 0.045\,755\,6(39) \, \meV \,,
\\
E^{(2)}(3 {}^3 D_2) =& \; \phantom{xx} 0.025\,460(4) \, \meV \,.
\end{align}
\end{subequations}
For the $4 {}^3 D_2$ state, the individual contributions are
\begin{subequations}
\begin{align}
E^{(2)}_{\rm K}(4 {}^3 D_2) =& \, -0.002\,567\,7 \, \meV \,,
\\
E^{(2)}_{\rm M}(4 {}^3 D_2) =& \; -0.005\,537\,9 \, \meV \,,
\\
E^{(2)}_\SO(4 {}^3 D_2) =& \; -0.001\,529\,0 \, \meV \,,
\\
E^{(2)}_{\FSS}(4 {}^3 D_2) =& \; \phantom{xx}\, 0.000\,925\,0 \, \meV \,,
\\
E^{(2)}_{\rm Q}(4 {}^3 D_2) =& \; \phantom{xx}\, 0.016\,254\,4(14) \, \meV \,,
\\
E^{(2)}(4 {}^3 D_2) =& \; \phantom{xx}\, 0.007\,544\,9(14) \, \meV \,.
\end{align}
\end{subequations}
Results for the total second-order terms $E^{(2)}$,
\begin{equation} \label{E2_total}
E^{(2)} = E^{(2)}_{\rm K} + E^{(2)}_{\rm M} + 
E^{(2)}_\SO + E^{(2)}_\FSS + E^{(2)}_{\rm Q}
\end{equation}
for the $3P$, $3D$, $4D$ and $4F$ states
of deuteronium, are given in Table~\ref{table4}.

\begin{table}[t!]
\sisetup{table-number-alignment = left,
  tight-spacing = true }
\caption{\label{table4}
Expectation values
$E^{(2)}$ of the second-order matrix elements [see Eq.~\eqref{E2_total}] are given
in meV for the $3P$, $3D$, $4D$ and $4F$ states of deuteronium.
The main uncertainty affecting the $E^{(2)}$ corrections is from 
the uncertainty in $Q_d$.}
\def\arraystretch{1.2}
\begin{tabular}{c@{\hspace*{0.3cm}} 
  S[table-format = 1.6e+2] @{\hspace*{0.3cm}} 
  c @{\hspace*{0.3cm}}
  S[table-format = 1.6e+2] }
\hline
\hline
$3{}^{2S+1}P_J$ & 
\multicolumn{1}{l}{$E^{(2)}$} & 
$3{}^{2S+1}D_J$ & 
\multicolumn{1}{l}{$E^{(2)}$} \\
\hline
$3^{-}\!P_1$ & -0.56801(5)    & $3^{-}\!D_2$ & -0.082692(7) \\
$3^{+}\!P_1$ &  0.85626(10) & $3^{+}\!D_2$ & 0.014097(5) \\
$3^{3}P_0$ & -1.12341(8)     & $3^{3}D_1$ & -0.079867(4) \\
$3^{3}P_1$ &  0.34352(4)     & $3^{3}D_2$ &  0.025460(4) \\
$3^{3}P_2$ &  -0.141812(8)  & $3^{3}D_3$ & -0.02238038(11) \\
$3^{5}P_2$ &  -0.72461(6)    & $3^{5}D_0$ &  0.041884(8) \\
$3^{5}P_3$ &  0.164931(15) & $3^{5}D_1$ &  0.003036(4) \\
                    &                        & $3^{5}D_3$ &  -0.067911(5) \\
                    &                        & $3^{5}D_4$ &   0.023279(2) \\
\hline
\hline
$4{}^{2S+1}D_J$ & 
\multicolumn{1}{l}{$E^{(2)}$} & 
$4{}^{2S+1}F_J$ & 
\multicolumn{1}{l}{$E^{(2)}$} \\
\hline
$4^{-}\!D_2$ & -0.029411(3)    & $4^{-}\!F_3$  & -0.0047254(4) \\
$4^{+}\!D_2$ &  0.002043(2)   & $4^{+}\!F_3$ & -0.0011946(3) \\
$4^{3}D_1$ & -0.0298719(14) & $4^{3}F_2$   & -0.0047244(2) \\
$4^{3}D_2$ &  0.0075449(14) & $4^{3}F_3$   &  0.0005208(2) \\
$4^{3}D_3$ & -0.0099559(4)   & $4^{3}F_4$   & -0.00195927(7) \\
$4^{5}D_0$ &  0.013379(3)     & $4^{5}F_1$   &  0.0000252(4) \\
$4^{5}D_1$ & -0.0004211(14) & $4^{5}F_2$   & -0.00224033(8) \\
$4^{5}D_3$ & -0.025625(2)     & $4^{5}F_4$   & -0.0038294(2) \\
$4^{5}D_4$ &  0.0067701(8)   & $4^{5}F_5$   &  0.00107371(14) \\
\hline
\hline
\end{tabular}
\end{table}

%
%
\section{Conclusions}
\label{sec7}

The calculation of the spectrum of 
heavy bound systems of
particles heavier than the electron
constitutes an involved task because
radiative and recoil corrections are
more intertwined than for lighter 
systems (for details, see Ref.~\cite{PaEtAl2024}).
In the heavy systems, electronic vacuum-polarization 
takes the role of the dominant 
radiative correction to energy levels,
being larger than self-energy effects
(see Refs.~\cite{Pa1996mu,Je2011pra,%
Bo2011preprintv7,Bo2012,VePa2004,KaIvKo2012,KaIvKa2013,KaKoShIv2017,AdJe2024gauge}
as well as Refs.~\cite{AdJe2025prl,AdJe2025prr}).

In our investigations, we have generalized
the calculation of relativistic-recoil
corrections to electronic vacuum polarization
to bound systems where the constituent particles
are either spinless or have spin $1/2$ or spin $1$.
To this end, we have used a version of NRQED where the mass scale separating 
soft from hard energies is much larger than the 
electron mass~\cite{AdJe2024gauge}.  This extension
of NRQED has been identified in Ref.~\cite{HiLePaSo2013}
and discussed in the context of muonic atoms in
Ref.~\cite{AdJe2024gauge}.
From the interaction kernels,
we have verified the Breit (interaction)
Hamiltonian for particles of 
spin-1, in Sec.~\ref{sec3}, and discussed the
generalization to spins 0 and 1/2.
The Breit Hamiltonian is based
on the exchange of massless photons [for
the propagator, see Eq.~\eqref{DC}].
The result is given in Eq.~\eqref{H4}.
For the treatment of vacuum polarization,
we have generalized the derivation of the
effective Hamiltonian (see Sec.~\ref{sec4})
to the exchange of a massive photon,
formulated in the one-loop optimized gauge [see
Eq.~\eqref{DOC1}].
We have derived the Hamiltonian for 
the massive exchange in Eq.~\eqref{calH4}
and have discussed its application to systems composed 
of particles having spins 0, 1/2, and 1.

The application of our formalism 
to bound systems of interest has been discussed in 
Sec.~\ref{sec5}, with a special emphasis
on  pionium, muonic hydrogen and muonic deuterium.
In Sec.~\ref{sec6}, a bound system consisting 
of two spin-$1$ particles.
is investigated, namely,
the bound system of a deuteron and 
its antiparticle, referred to as 
deuteronium. 
The results reported here 
remove an obstacle in the process of advancing
the theory of the bound-state spectrum
of deuteronium~\cite{AdJe2025prl,AdJe2025prr}
to the $\alpha^5 m_d$ order. Furthermore,
in the context of heavy bound 
systems, the full consideration of the 
$\alpha^5 m_r$ corrections, for 
the bound-state spectra of bound
systems with spins $0$, $1/2$ and $1$,
enables sensitive tests for possible
low-energy extensions of the Standard 
Model~\cite{KrEtAl2016,KrEtAl2017woc1,%
KrEtAl2017woc2,KrEtAl2019,AlEtAl2023,KrEtAl2024}.

\section*{Acknowledgments}

This work was supported by the National Science Foundation through Grants
PHY-2308792 (G.S.A.) and PHY--2513220 (U.D.J.), and by the National Institute
of Standards and Technology Grant 60NANB23D230 (G.S.A.).

\appendix

%
%
\section{Fourier Transforms}
\label{appa}

Some useful three-dimensional Fourier transforms are
\begin{subequations} \label{Fourier} 
\begin{align}
\int \frac{\dd^3 k}{(2\pi)^3} \, 
\ee^{\ii \vec k \cdot \vec x} 
\frac{1}{\vec k\,^2+ \lambda^2} =&  \;
\frac{\ee^{-\lambda r}}{4 \pi r} \, , \\[3pt]
\int \frac{\dd^3 k}{(2\pi)^3} \, 
\frac{\vec k\,^2 \, 
\ee^{\ii \vec k \cdot \vec x} }{\vec k\,^2+ \lambda^2} =&  \;
\delta^{(3)}(\vec x\,) - \lambda^2 \, \frac{\ee^{-\lambda r}}{4 \pi r} \, , 
\\[3pt]
\int \frac{\dd^3 k}{(2\pi)^3}  \, 
\frac{\lambda^2 \ee^{\ii \vec k \cdot \vec x}}%
{\vec k\,^2 (\vec k\,^2+\lambda^2)} =&  \;
\frac{\ee^{-\lambda r}}{4 \pi r} \left ( \ee^{\lambda r}-1 \right )  \,.
\end{align}

For integrals with the square of $( \vec k\,^2+ \lambda^2 )$
in the denominator, we obtain
\begin{align}
\int \frac{\dd^3 k}{(2\pi)^3}  \,
\frac{\ee^{\ii \vec k \cdot \vec x}}{( \vec k\,^2+ \lambda^2 )^2} 
=& \; \frac{\ee^{-\lambda r}}{8 \pi \lambda} \,, \\
\int \frac{\dd^3 k}{(2\pi)^3} \,
\frac{\vec k\,^2 \, \ee^{\ii \vec k \cdot \vec x}}%
{( \vec k\,^2+ \lambda^2 )^2} =&  \;
\frac{\ee^{-\lambda r}}{4 \pi r} 
\left ( 1 - \frac{\lambda r}{2} \right ) \,, 
\\[3pt]
\int \frac{\dd^3 k}{(2\pi)^3} \, 
\frac{\lambda^4 \ee^{\ii \vec k \cdot \vec x}}%
{\vec k\,^2 (\vec k\,^2+\lambda^2)^2} =&  \;
\frac{\ee^{-\lambda r}}{4 \pi r} 
\left ( \ee^{\lambda r} - 1 - 
\tfrac{\lambda r}{2} \right ) \, .
\end{align}
\end{subequations} 
The following vector-valued integrals are also relevant,
\begin{subequations} 
\begin{align}
\int \frac{\dd^3 k}{(2\pi)^3} \, 
\ee^{\ii \vec k \cdot \vec x} 
\frac{ k^i}{\vec k\,^2+ \lambda^2} =&  \;
\frac{\ee^{-\lambda r}}{4 \pi r^2} (1+\lambda r) \, \ii \, 
\hat x^i \, ,  \\[3pt]
\int \frac{\dd^3 k}{(2\pi)^3}  \, 
\frac{\lambda^2 k^i \ee^{\ii \vec k \cdot \vec x}}%
{\vec k\,^2 (\vec k\,^2+\lambda^2)} =&  \;
\frac{\ee^{-\lambda r}}{4 \pi r^2} 
\left ( \ee^{\lambda r}-1- \lambda r \right ) \, \ii \, \hat x^i \,, \\[3pt]
\int \frac{\dd^3 k}{(2\pi)^3}  \, 
\ee^{\ii \vec k \cdot \vec x} \frac{k^i}{( \vec k\,^2+ \lambda^2 )^2} =&  \;
\frac{\ee^{-\lambda r}}{8 \pi} \, \ii \, \hat x^i \, , 
\end{align}
\end{subequations} 
In these calculations, the general identity for Fourier 
transforms 
${\FTK} \Big [ k^i f(\vec k\,) \Big](\vec x\,) = 
-\ii \partial_i {\FTK} \Big [ f(\vec k\,) \Big ](\vec x\,)$ was used,
along with partial-fraction decompositions such as 
\begin{equation}
\frac{1}{\vec k\,^2+\lambda^2} = \frac{1}{\vec k\,^2} -
\frac{\lambda^2}{\vec k\,^2 (\vec k\,^2+\lambda^2)}.
\end{equation}
The relevant integrals of second-rank tensors are
\begin{align}
\int \frac{\dd^3 k}{(2\pi)^3} \, \ee^{\ii \vec k \cdot \vec x} \,
\frac{k^i k^j}{\vec k\,^2} =& \;
\frac{1}{3} \, \delta^{i j} \, \delta^{(3)} (\vec x\,) 
\\
& \; + \frac{1}{4 \pi r^3} \left ( \delta^{i j} - 
3 \hat x^i \hat x^j \right ) \, , 
\nonumber\\[3pt]
\int \frac{\dd^3 k}{(2\pi)^3} \, 
\ee^{\ii \vec k \cdot \vec x} \,
\frac{k^i k^j}{\vec k\,^4} =&  \;
\frac{1}{8 \pi r} \left ( \delta^{i j} - \hat x^i \hat x^j \right ) \, , \\[3pt]
\int \frac{\dd^3 k}{(2\pi)^3} \, 
\ee^{\ii \vec k \cdot \vec x} 
\frac{k^i k^j}{( \vec k\,^2+ \lambda^2 )^2} =&  \;
\frac{\ee^{-\lambda r}}{8 \pi r} 
\big ( \delta^{i j} - (1+\lambda r) \, \hat x^i \hat x^j \big ) \,.
\end{align}
The ones here without $\lambda$ factors were evaluated as the $D \rightarrow 3$
limits of the general-$D$ results \cite{Ad2016BALMS}.
Two further tensor integrals have more complicated expressions.
The first of these is
\begin{multline}
\int \frac{\dd^3 k}{(2\pi)^3} \, 
\ee^{\ii \vec k \cdot \vec x} \frac{k^i k^j}{\vec k\,^2+\lambda^2} =
\frac{1}{3} \, \delta^{i j} \, \delta^{(3)} (\vec r) 
\\
+ \frac{\ee^{-\lambda r}}{4 \pi r^3} 
\Big \{ \left ( 1+\lambda r \right ) \delta^{i j} -
\left ( 3 + 3 \lambda r + \lambda^2 r^2 \right ) \, 
\hat x^i \hat x^j \Big \} \, , 
\end{multline}
while the second is
\begin{multline}
\int \frac{\dd^3 k}{(2\pi)^3}  \, 
\ee^{\ii \vec k \cdot \vec x} \,
\frac{\lambda^2 k^i k^j}{\vec k\,^2 (\vec k\,^2+\lambda^2)} 
= \frac{\ee^{-\lambda r}}{4 \pi r^3} 
\biggl\{ \left ( \ee^{\lambda r} -1-\lambda r \right ) \delta^{i j} 
\\
- 3 \left ( \ee^{\lambda r}-1- \lambda r-\frac{\lambda^2 r^2}{3} \right ) \, 
\hat x^i \hat x^j \biggr\} \,.
\end{multline}

\end{document}